\documentclass[5p,times]{elsarticle}

\usepackage{amssymb}

\journal{Future Generation Computer Systems}

\usepackage{xcolor}
\usepackage{url}
\usepackage{listings}
\usepackage{graphicx}
\usepackage{subfig}
\usepackage[USenglish]{babel}
\usepackage{framed}
\usepackage{multirow}
\usepackage{colortbl}
\usepackage{enumitem}
\usepackage{array}
\usepackage{comment}

\newcolumntype{P}[1]{>{\centering\arraybackslash}p{#1}}
\newcolumntype{M}[1]{>{\raggedright\arraybackslash}m{#1}}
\newcolumntype{C}[1]{>{\centering\arraybackslash}m{#1}}

\newcommand{\ihpc}{{\fontfamily{cmss}\selectfont IgnisHPC}}

\definecolor{lightgray}{gray}{0.9}
\definecolor{shadecolor}{gray}{0.92}
\definecolor{dkgreen}{rgb}{0,0.6,0}
\definecolor{gray}{rgb}{0.5,0.5,0.5}
\definecolor{mauve}{rgb}{0.58,0,0.82}

\begin{document}

\begin{frontmatter}



\title{A unified framework to improve the interoperability between HPC and Big Data languages and programming models\tnoteref{t1}}
\tnotetext[t1]{This work has been supported by MICINN (RTI2018-093336-B-C21), Xunta de Galicia (ED431G-2019/04 and ED431C-2018/19) and the European Regional Development Fund (ERDF).}
	

\author[1]{César Piñeiro\corref{cor1}}
\ead{cesaralfredo.pineiro@usc.es}
\author[1]{Juan C. Pichel}
\ead{juancarlos.pichel@usc.es}

\affiliation[1]{organization={Centro Singular de Investigaci\unexpanded{ó}n en Tecnolox\unexpanded{í}as Intelixentes (CiTIUS), Universidade de Santiago de Compostela},
            city={Santiago de Compostela},
            postcode={15782}, 
            country={Spain}}
\cortext[cor1]{Corresponding author}

\begin{abstract}
One of the most important issues in the path to the convergence of HPC and Big Data is caused by the differences in their software stacks. Despite some research efforts, the interoperability between their programming models and languages is still limited. To deal with this problem we introduce a new computing framework called \ihpc{}, whose main objective is to unify the execution of Big Data and HPC workloads in the same framework. \ihpc{} has native support for multi-language applications using JVM and non-JVM-based languages. Since MPI was used as its backbone technology, \ihpc{} takes advantage of many communication models and network architectures. Moreover, MPI applications can be directly executed in a efficient way in the framework. The main consequence is that users could combine in the same multi-language code HPC tasks (using MPI) with Big Data tasks (using MapReduce operations). The experimental evaluation demonstrates the benefits of our proposal in terms of performance and productivity with respect to other frameworks such as Apache Spark. \ihpc{} is publicly available for the Big Data and HPC research community.

\end{abstract}



\begin{keyword}
Big Data \sep HPC \sep MPI \sep Multi-language \sep Programming models 


\end{keyword}

\end{frontmatter}


\section{Introduction}
\label{sec:intro}

The unification of High Performance Computing (HPC) and Big Data has received increasing attention in the last years. It is a common belief that exascale computing and Big Data are closely associated since HPC requires processing large-scale data from scientific instruments and simulations. But, at the same time, it was observed that tools and cultures of HPC and Big Data communities differ significantly~\cite{Hel20}. One of the most important sources of divergence comes from the differences between their software ecosystems. In this way, HPC applications have traditionally been based on MPI (Message Passing Interface) to support inter-node parallel execution, and based on OpenMP or other alternatives to exploit intra-node parallelism. However, Big Data programming models are based on interfaces like Hadoop~\cite{Whi15} or Spark~\cite{Zah10}. In addition to different programming models, programming languages also differ between both communities: being Fortran and C/C++ the most common languages in HPC applications, and Java, Scala, or Python being the most common languages in Big Data applications. 

This divergence between programming models and languages sets out a convergence problem, not only related to the interoperability of the applications but also to the interoperability between data formats from different programming languages~\cite{Asc18}. In this scenario, we need to consider how to build end-to-end workflows where, for example, simulations can be MPI applications written in Fortran or C/C++, and the analytics codes can be written in Java or Python (maybe parallelized by a Big Data framework). 

In this work we introduce a new computing framework called \ihpc{}\footnote{It is publicly available at \url{https://github.com/ignishpc}} to deal with that issue. The main goal of \ihpc{} is to unify in the same framework the development, combination and execution of HPC and Big Data applications using different languages and programming models. With this objective in mind, we can summarize the main contributions of \ihpc{} as follows:
\begin{itemize}[noitemsep]
    \item Unlike other frameworks such as Hadoop and Spark, \ihpc{} supports natively both JVM and non-JVM-based languages. Applications can be implemented using one or several programming languages following an API inspired by Spark's one.  
    \item \ihpc{} uses MPI as backbone technology, which allows the framework to support many communication models and network architectures. In addition, MPI applications and libraries can be directly executed in an efficient way in \ihpc{}. In this way, most of the HPC scientific applications, which in many cases contain tens of thousands of lines of code, do not have to be ported to a new API or programming model. Nowadays, to the best of our knowledge, there is no other Big Data framework with that feature.
    \item In \ihpc{}, MPI codes can be easily combined with typical MapReduce operations to create hybrid applications. Therefore, users can use the programming models that best fit their data-intensive and compute-intensive tasks in the same application.
    \item A thorough experimental evaluation has been carried out to demonstrate the benefits of \ihpc{} in terms of performance and productivity. The study showed that \ihpc{} clearly outperforms Spark when considering different types of Big Data application patterns. Moreover, we proved that running MPI (and hybrid MPI+OpenMP) applications from \ihpc{} is easy and as efficient as  executing them natively.
    \item To avoid dependencies, \ihpc{} is fully containerized and supports some of the  most well-known resource and scheduler managers. 
\end{itemize}

The remainder of this paper is organized as follows. Section \ref{sec:background} provides some context about Big Data and HPC technologies. Section \ref{sec:ignishpc} explains in detail the architecture and modules of \ihpc{}. Section \ref{sec:programming} describes how to implement applications using the \ihpc{} API. Section \ref{sec:integrating_mpi} focuses on the integration of MPI in \ihpc{}, and how MPI applications can be executed within the framework. The experimental evaluation is shown in Section \ref{sec:exp_eval}. Section \ref{sec:related} discusses the related work. Finally, the main conclusions derived from this work are explained.

\section{Background}
\label{sec:background}

\subsection{Big Data frameworks}
\label{sec:bd_framewroks}

MapReduce is a programming model introduced by Google for processing and generating large data sets on a huge number of computing nodes~\cite{dea04}. Apache Hadoop~\cite{Whi15} was the first open-source implementation of the MapReduce programming model. It was widely adopted by both industry and academia, thanks to that simple yet powerful programming model that hides the complexity of parallel task execution and fault-tolerance from the users. However, most applications do not fit this model and require a more general data orchestration.

Apache Spark~\cite{Zah10} was designed to overcome some of the Hadoop limitations, especially when considering iterative jobs. Nowadays Spark is considered the \emph{de-facto} standard for Big Data processing. Unlike Hadoop, Spark uses Resilient Distributed Datasets (RDDs) which implement in-memory data structures used to cache intermediate data across a set of nodes. Since RDDs can be kept in memory, algorithms can iterate over RDD data many times very efficiently. In addition, Spark provides many attractive features such as fault-tolerance, as RDDs can be regenerated through lineage when compute nodes are lost. Spark uses a thread-based worker model for executing the tasks. In this way, a Spark job is controlled by a driver program, which usually runs in a separate master node. On the other hand, the parallel regions in the driver program are shipped to the cluster to be executed. 

Spark is implemented in Scala and also has interfaces to execute Java, Python and R applications. Both Hadoop and Spark are capable of running codes written in other programming languages, but they suffer performance issues since they require sharing data outside the Java Virtual Machine (JVM) through system pipes~\cite{Ding2011}. Contrary to the common belief, this behavior also applies to Python in Spark because, while the Python driver code can be executed within the JVM thanks to Jython~\cite{Jython}, executors are directly executed with the available Python interpreter. As a consequence, the performance of Python codes is affected like any non-JVM language such as C++~.

In our previous work~\cite{Pin20}, we introduced Ignis, a first attempt to build an efficient and scalable multi-language Big Data framework. Despite being a prototype, Ignis has two significant contributions with respect to Spark. First, it allows to execute natively applications implemented using non-JVM languages such as Python and C/C++. Second, it supports multi-language applications, so different computing tasks could be implemented in the programming language that best suits them. However, there are also some important limitations as the following. For instance, Ignis is restricted to use TCP sockets for inter-node communication, so it is not possible to take advantage of typical HPC networks such as Infiniband, Myrinet, etc. Another issue is that Ignis only supports one data partition per worker (executor). As a consequence, in order to work with big datasets, it is necessary to create new workers, which causes a degradation in the I/O performance. Ignis is completely containerized, but it uses a very limited ad-hoc solution for containers orchestration (named \emph{Ancoris}). And finally, there was no submitter in Ignis, so jobs are manually launched using several configuration scripts. 

\subsection{MPI}

Message Passing Interface (MPI) is the most widely used and dominant programming model in HPC. In MPI, processes make explicit calls to library routines defined by the MPI standard to communicate data between two or more processes. These routines include both point-to-point (two party) and collective (many party) communication. Note that from MPI-3 new features were introduced to enable MPI processes within an SMP node to collectively allocate shared memory for direct load and store operations, which enables the shared-memory-processes to more efficiently share data.

Among the different MPI implementations, the most successful ones are MPICH~\cite{mpich} and Open-MPI~\cite{openmpi}. In particular, \ihpc{} uses MPICH as backbone technology to perform all the communications between processes, which allows the framework to support many communication models and network architectures (e.g. Infiniband, Myrinet, etc.). As we will explain later, one of the most important consequences of this design decision is that \ihpc{} can efficiently execute native MPI applications. Therefore, it is not necessary to port MPI applications using a different API. In this way, \ihpc{} is able to bring together the benefits of HPC and Big Data worlds into the same framework. 

We have considered MPICH instead of other MPI implementations because it is a mature project and provides several important features that are critical in a Big Data environment. For example, it is possible to join processes dynamically. It means that MPICH allows to connect independent instances of a MPI process at runtime, which is essential for increasing the number of processes when more parallelism is needed or for replacing lost processes when a computing node fails. 

\subsection{Resource managers and schedulers}

In a Big Data environment, it is necessary to manage and balance the cluster resources to allow multiple applications and frameworks to be efficiently executed together on the same system. That is the goal of resource managers such as Apache Mesos~\cite{Hin11} and Nomad~\cite{nomad}. 

In particular, Apache Mesos groups all the physical resources of each node of the cluster and make them available for the applications as a single pool of resources. Among the features of Mesos we can highlight that it provides resource isolation thanks to its support to Docker containers~\cite{mer14}. So it allows the execution of jobs in a custom independent environment both in terms of resources and installed software. 

Users cannot interact directly with Mesos because it is only a resource planner, so an orchestration framework is required to run and schedule tasks. We can find many orchestration frameworks depending on the type of tasks to be executed. In this work we have considered two of the most relevant, Apache Marathon~\cite{marathon} and Apache Singularity~\cite{singularity}, which both support Docker containers orchestration.

Finally, Nomad is a simple workload orchestrator created by HashiCorp. Its flexible and consolidated workflow provides users the functionality of both a resource and a scheduler manager, combined into a single system. Nomad is a modern alternative to Mesos and also supports containerized and non-containerized applications with different life cycle. Unlike old legacy platforms, it is prepared for running heterogeneous applications and using accelerators such as GPUs in a simple way. 

\begin{figure}[t!]
 \centering
 \subfloat{\includegraphics[width=0.3\textwidth]{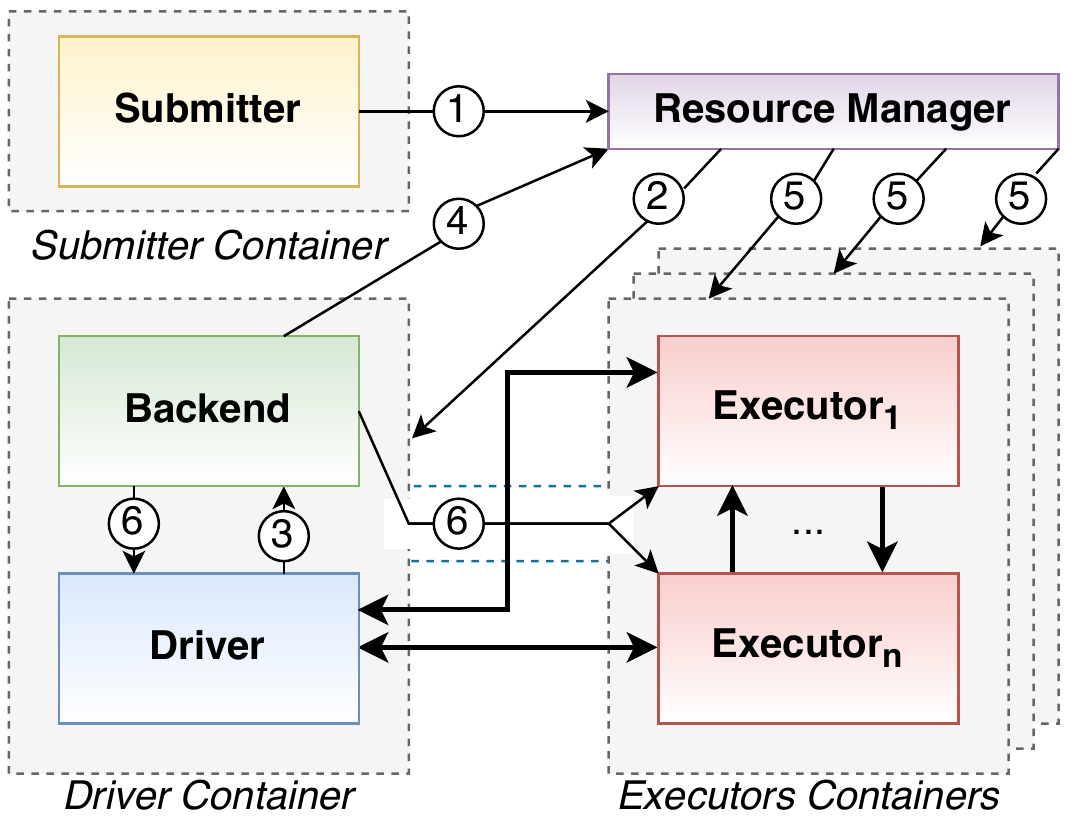}}
 \caption{Scheme of the architecture of \ihpc{}.}
 \label{fig:architecture} 
\end{figure}

\section{\ihpc{}}
\label{sec:ignishpc}

\subsection{Architecture of the framework}

We can divide the \ihpc{} architecture into four independent modules which run inside Docker containers: \textsc{Submitter}, \textsc{Backend},  \textsc{Driver} and \textsc{Executor}. A scheme is shown in Figure \ref{fig:architecture}. Modules were implemented in different languages, using Apache Thrift\footnote{https://thrift.apache.org} for the inter-module communications. In the figure, thin arrows represent those RPC communications, bold arrows data transfers, and numbers indicate the call order. 

The \textsc{Submitter} module is in charge of launching the \ihpc{} jobs (1) making a request to the resource manager (and scheduler), which is an external dependency that is responsible of the containers orchestration and the management of their resources. Afterwards, the  driver container is started (2), where the \textsc{Driver} module is the container entrypoint. That module exposes all the available features of \ihpc{} through a user API created as interface to the \textsc{Backend}, where the API logic is implemented as a service (3).
The \textsc{Backend} module is started inside the driver container after the driver code initializes the framework, and it is responsible  of making the requests to the resource manager following the instructions specified in the driver code (4). As a consequence, the resource manager will create the executors containers (5). The \textsc{Executor} module contains the low-level implementation of a set of operations required by the \textsc{Backend} for each supported programming language. Note that \ihpc{} uses a SSH tunnel to connect driver and executors containers to handle executors in a safe and secure manner. Finally, the \textsc{Backend} is connected to the executors in order to perform the low-level API operations (6). It is important to highlight that the \textsc{Driver} is also considered an executor by the \textsc{Backend} to handle the data transfers. 

There are important architectural differences between \ihpc{} and Ignis~\cite{Pin20}, our first prototype of a multi-language Big Data framework. Changes performed to \ihpc{} focused on removing some important limitations and performance issues shown by Ignis (see Section \ref{sec:bd_framewroks}). Among them we can highlight the following:
\begin{itemize}[noitemsep]
    \item One of the main goals of \ihpc{} is to unify the execution of Big Data and HPC workloads in the same framework. For this reason data transfers in \ihpc{} (bold arrows in Figure \ref{fig:architecture}) are performed using MPI. It has enormous advantages over the inter-node communications with TCP sockets used by Ignis. First, \ihpc{} supports many different communication models and network architectures (e.g. Infiniband,  Myrinet, etc.). In this way, it covers the characteristics of the vast majority of Big Data and/or HPC clusters. Moreover, MPI applications and libraries can be directly executed in \ihpc{}. It means that HPC scientific applications, which in many cases contain tens of thousands of lines of code, do not have to be ported to the \ihpc{} API. And finally, it is possible to combine in the same multi-language code HPC tasks (using MPI) with Big Data tasks (using MapReduce operations).    
    \item \ihpc{} has a new \textsc{Submitter} module that handle jobs using an external resource manager. This module includes a submit script, similar to Spark's \texttt{spark-submit}, that allows users to easily launch \ihpc{} jobs. On the contrary, jobs in Ignis are manually configured and launched using several scripts. There is no \textsc{Submitter} module in its architecture. 
    \item Ignis used a \textsc{Manager} module (one per executor container) that acted as middleman between the \textsc{Backend} module and the executors. To be more efficient, \ihpc{} removed that module and its functionalities were adopted by the \textsc{Backend}. 
    \item \ihpc{} supports some of the most well-known resource and scheduler managers. In addition, it was designed to easily add new managers. However, Ignis is bonded to work only with an ad-hoc manager, which has limited functionalities. 
\end{itemize}

\subsection{Jobs in the framework}

\begin{figure}[t!]
 \centering
 \includegraphics[width=0.25\textwidth]{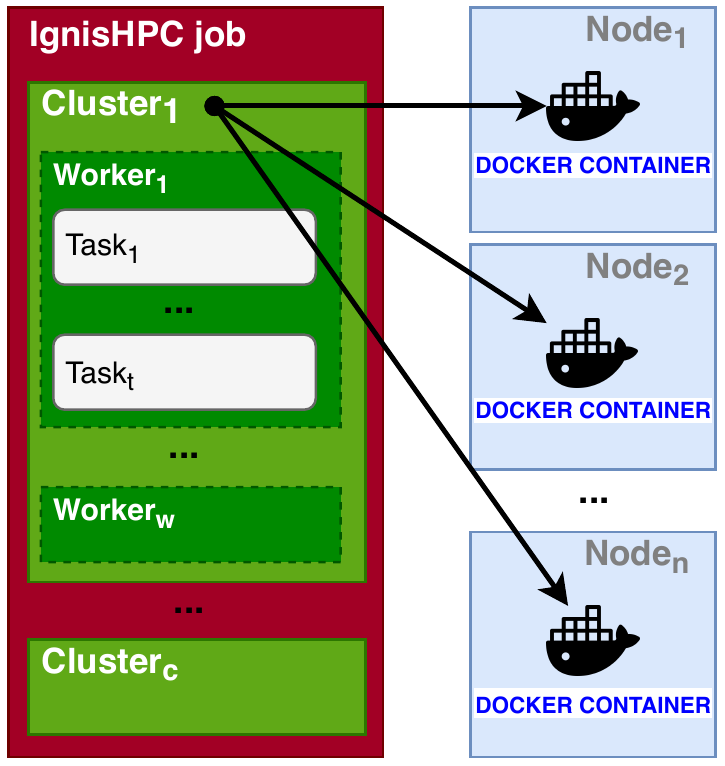}
 \caption{Job hierarchy in the \ihpc{} framework.}
 \label{fig:classes} 
\end{figure}

It is important to know the structure of an \ihpc{} job. It consists of a set of Docker containers distributed in multiple computing nodes grouped in \emph{Clusters} (see Figure~\ref{fig:classes}). \emph{Workers} are bonded to a single programming language and run inside a \emph{Cluster}, so at least one \emph{Worker} has to be created for each programming language in order to build multi-language applications. A \emph{Worker} instantiates at least one process (executor) on each Docker container with the aim of executing its tasks in parallel, processing them as a pipeline.

\emph{Clusters} are independent, each one has its own assigned resources, so they can execute different tasks at the same time. Using multiple \emph{Clusters} could be useful if stages or phases of the job have some compatibility issues. In this way, incompatible phases would be executed by different configured \emph{Clusters}. However, \emph{Workers} can be executed in shared mode (disabled by default). It means that executors of two or more \emph{Workers}, which are located in the same container, would share the available resources. Normally users configure each \emph{Worker} to use a part of the \emph{Cluster} resources (e.g. cores, memory, etc.).

\subsection{Resource manager}
\label{sec:resource}

Since \ihpc{} must be executed inside Docker containers, a resource and scheduler manager is required to handle the cluster resources and launch these containers. Note that any framework that meets these requirements could be used in \ihpc{} by implementing a basic interface. Currently \ihpc{} supports the following managers:

\begin{itemize}[noitemsep]
    \item \emph{Docker}: It is the easiest way to run \ihpc{} locally on a single machine. It uses the Docker client to directly launch containers.    
    \item \emph{Ancoris}~\cite{Pin20}: This is the only one supported by Ignis and it was designed as a simple and light resource manager. It executes itself inside Docker containers and is composed of two types of instances: master and slaves. The master manages the available resources in the cluster and it is responsible of launching the containers. Slaves expose the resources of their host to the master when they are deployed.
    \item \emph{Mesos+Marathon}: Since Spark supports Mesos, using it as resource manager together with Marathon as container orchestrator allows users to execute in a single environment Spark and \ihpc{} jobs. In addition, since \ihpc{} is able to execute efficiently Big Data and MPI-based applications, we are merging both Big Data and HPC software ecosystems in just one execution environment. 
    \item \emph{Mesos+Singularity}: The same benefits commented above apply to the combination of Mesos and Singularity. 
    \item \emph{Nomad}: It combines in the same framework a resource and a scheduler manager. Due to its lack of dependencies, it is the best option to install in a cluster from scratch. Moreover, it has better support for devices like GPUs than Mesos, which allows to create heterogeneous execution environments. 
\end{itemize}

\subsection{Driver module}

The \textsc{Driver} module is a user API that allows access to all \ihpc{} functionalities. The driver program describes the high-level control flow of the application, and it can be programmed in any of the supported languages (currently, Java, Python and C/C++). The \textsc{Driver} was designed as a Thrift RPC interface to the \textsc{Backend} so the logic has not to be re-implemented for every programming language. More details about the driver API and how to implement an application in \ihpc{} are provided in Section \ref{sec:programming}.

\subsection{Backend module}

The \textsc{Backend} module contains the services that define the \textsc{Driver}'s logic. For instance, the \texttt{reduceByKey} function requires searching and grouping the keys. These operations are defined in the \textsc{Backend}, but they are implemented in the \textsc{Executor} module for a specific programming language. 

\begin{figure}[t!]
 \centering
 \includegraphics[width=0.3\textwidth]{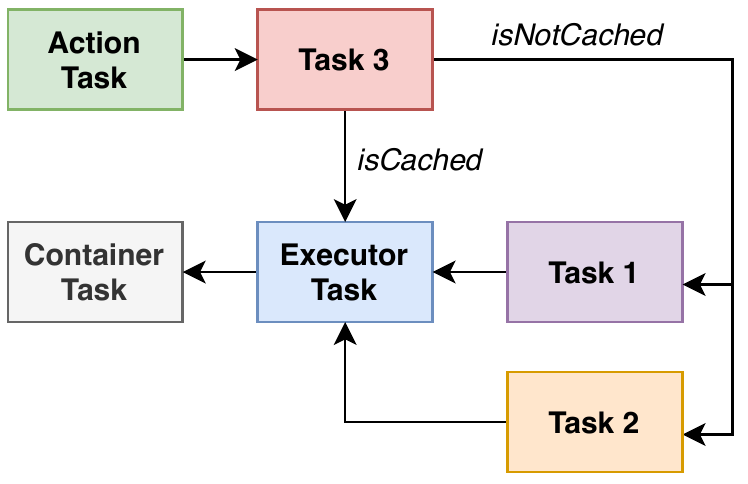}
 \caption{Example of a task dependency graph.}
 \label{fig:task_dependency} 
\end{figure}

The \textsc{Backend} module is also responsible of sending requests to the resource manager in accordance with the instructions specified in the driver code. These instructions are lazily executed, so the \textsc{Backend} registers the function calls as a task dependency graph. When a task that represent an action in the driver code is created (e.g. a call to \texttt{count}), all the tasks in its dependency graph are executed. An example is shown in Figure \ref{fig:task_dependency}, where the \emph{Action Task} depends on \emph{Task 3}, and \emph{Task 3} depends on \emph{Task 1} and \emph{2}. Note that a task dependency is only computed if the task was never executed or if its result was not explicitly cached. The \emph{Executor} and \emph{Container} tasks are always executed as final dependencies. These tasks check that executors and containers are running and ready to be used.  

Finally, \ihpc{} is able to recover after a failure of a cluster node or some of the executors. Affected tasks are traced by the \textsc{Backend} in such a way that only their executors are reallocated and recomputed. If the affected tasks are cached, the recovery process will be faster since it is not necessary to recalculate their dependencies. Note that this process is automatic, but users can tune the recovery process using the persistence functions in the driver code (see Section \ref{sec:programming}).

\subsection{Executor module}
\label{sec:executor}

The \textsc{Executor} module implements the operations defined by the \textsc{Backend}, where each supported programming language has its own implementation. In order to add support for a new language in \ihpc{}, a minimum implementation only requires programming the context class. The executor context allows the API functions to interact with the rest of the \ihpc{} system. In this way, among the functionalities of the context we find the exchange of user variables between driver and executors or providing the executors access to the MPI communicators. 

\begin{figure}[t!]
 \centering
 \includegraphics[width=0.35\textwidth]{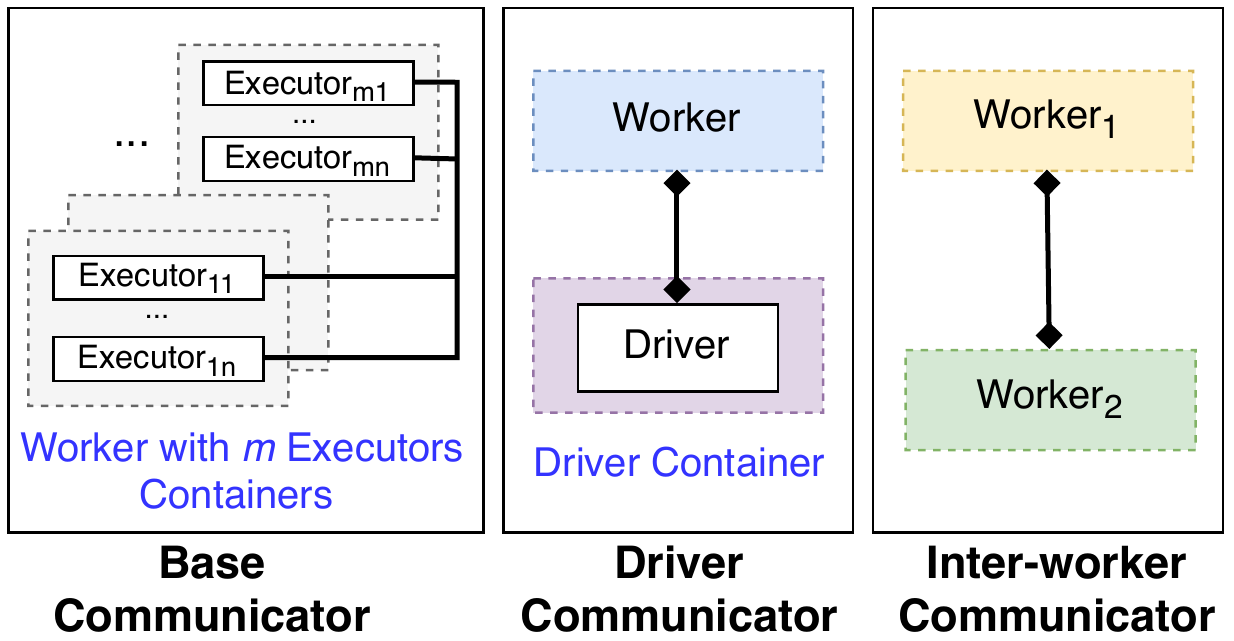}
 \caption{MPI communicators in \ihpc{}.}
 \label{fig:mpi_communicators} 
\end{figure}

As we explained previously, \ihpc{} uses MPI (that is, MPI communicators) to perform all the communications related to the \textsc{Executor} module. \ihpc{} constructs three types of communicators for data transfers (see Figure \ref{fig:mpi_communicators}): 

\begin{itemize}[noitemsep]
    \item \emph{Base communicator}: for each worker there is a communicator that includes all its executors. This communicator always exists. If one executor is lost, the communicator is destroyed and a new communicator is created including a new executor. To that end, the capability of linking dynamically a single process to a communicator introduced in MPI-3 was of special importance. Without this feature all processes would have to be launched at the same time, and in case a process died, it could not be replaced causing the job to fail.
    \item \emph{Driver communicator}: it joins a base communicator to the driver process. It is created when the driver and a worker exchange data.
    \item \emph{Inter-worker communicator}: it is created joining the base communicators of two workers. It is used to send data from one worker to another and it will be destroyed as soon as one of the two workers stops its execution. This communicator is only created between workers that execute the operation \texttt{ImportData}. 
\end{itemize}

The base communicator for each worker is accessible to programmers by the executor context. It means that \ihpc{} functions can be implemented using that communicator and MPI primitives (e.g. gather, scatter, broadcast, reduce, etc.). As a result, \ihpc{} supports the execution of pure MPI applications with minimal modifications in the original code. A detailed explanation is provided in Section \ref{sec:integrating_mpi}. 

Another benefit of using MPI for data transfers is the performance improvement of iterative applications. When using Big Data frameworks such as Ignis and Spark, an iterative application requires the driver to perform an evaluation task per iteration to obtain the final result. Each evaluation has three steps: stopping the executors, analysis of the partial results by the driver and restarting the executors. Note that starting and stopping the executors is very costly in terms of performance. \ihpc{} avoids the driver evaluations because executors share the partial results of each iteration using their MPI base communicator. Therefore, it is not necessary to stop them because they do not need to wait for the driver. This has even a bigger impact on performance for those applications with many short iterations.

\subsection{Submitter module}

The \textsc{Submitter} is an \ihpc{} module consisting of a set of scripts and utilities for configuring and launching jobs. As we commented previously, Ignis had no module to launch tasks, and the \textsc{Driver} module was launched manually using \emph{Ancoris} (see Section \ref{sec:resource}). The \textsc{Submitter} is a container, which can be accessed by ssh. There users can set up jobs in a similar environment where the \ihpc{} applications will be executed.

The main utility of the \textsc{Submitter} module is the \texttt{ig\-nis\--submit} script that, like \texttt{spark-submit}, allows users to launch \ihpc{} jobs in the cluster. The script only requires as mandatory arguments a Docker image and the driver program. There are also the following optional parameters: 

\begin{itemize}[noitemsep]
    \item \texttt{name}: a job name can be specified. 
    
    \item \texttt{arguments}: after the driver program name, all the parameters will be considered as driver arguments. 
    
    \item \texttt{attach mode}: by default, jobs are launched in unattached mode. That is, \texttt{ignis-submit} launches the job and exits. Attach mode allows users to control the job as if the driver runs locally, so output is printed in real time and it is possible to manually kill the job.  

    \item \texttt{properties}: users can change the default properties before launching the \textsc{Driver} module. \textsc{Executor} properties can be redefined later but \textsc{Driver} properties are set only by \texttt{ignis-submit}.
    
\end{itemize}

\begin{figure}[t]
\begin{lstlisting}[language=bash]
 # Python driver
 ignis-submit ignishpc/python python3 mydriver.py

 # C++ driver
 ignis-submit --name myapp --properties ignis.driver.memory=2GB ignishpc/cpp ./mydriver 0 -g 2
\end{lstlisting}
	\vspace{-0.35cm}
	\caption{Job submission examples.}
	\label{fig:submit_examples}    
\end{figure}

Figure \ref{fig:submit_examples} shows two job submission examples. The first one is a Python basic submission with only its base image and the driver application. The second one deals with a C++ driver and optional parameters. In particular, \texttt{--name} sets the job name, \texttt{--properties} changes the driver default memory to 2 GB,  and \texttt{0 -g 2} are considered arguments of \texttt{mydriver}. Note that \texttt{ignishpc/cpp} is the C++ base image.

\subsection{Data storage}
\label{sec:data_storage}

\ihpc{} provides multiple options for data storage. Users can choose a type of storage according to their particular execution environment. Storage must be defined as a property before the worker creation. \ihpc{} supports the following storage options: 

\begin{itemize}[noitemsep]
    \item \emph{In-Memory:} it is the best performer since all data is stored in memory. Memory consumption could be an issue so it is not suitable for all kinds of jobs. 
    
    \item \emph{Raw memory:} data is stored in a memory buffer using a serialized binary format. Extra memory consumption is minimal and the buffer is compressed by Zlib~\cite{Gopal2014}, which has nine compression levels. Level six is applied by default, but it can be changed when the worker properties are defined.
    
    \item \emph{Disk:} similar to raw memory but the buffer is implemented as a file. Performance is much lower but it allows to work with large amounts of data that cannot be completely stored in memory.
\end{itemize}

On the other hand, there is an important difference in how memory is handled by Ignis and \ihpc{}. Since Ignis was just a prototype, for simplicity in the implementation, it assigns one data partition to each executor. In this way, if it is necessary to increase the partition size, a realloc operation is performed in such a way that the complete partition is copied to a different memory location. The consequence is a noticeable increment in the memory consumption. This restricts Ignis to work with smaller input datasets. \ihpc{} overcomes that limitation supporting several data partitions per executor. Note that an executor can spawn several threads to process the data partitions in parallel.

\begin{table}[t!]
\footnotesize
\centering
\begin{tabular}{M{2cm}M{5cm}}
\rowcolor{lightgray}
\hline
{\bf Type}  & {\bf Functions} \\ \hline
Conversion  &    \texttt{map, filter, flatmap, keyBy, mapPartitions, keys, values, mapValues,} etc.\\ \hline
Group       &    \texttt{groupBy, groupByKey} \\ \hline
Sort        &    \texttt{sort, sortBy, sortByKey} \\ \hline
Reduce      &    \texttt{reduce, treeReduce, aggregate, treeAggregate, fold, reduceByKey, aggregateByKey,} etc. \\ \hline
I/O          &    \texttt{collect, top, take, saveAsObjectFile, saveAsTextFile, saveAsJsonFile,} etc\\ \hline
SQL         &    \texttt{union, join, distinct} \\ \hline
Math       &    \texttt{sample, sampleByKey, takeSample, count, max, min, countByKey, countByValue} \\ \hline
Balancing   &    \texttt{repartition, partitionBy} \\ \hline
Persistence &    \texttt{persist, cache, unpersist, uncache} \\ \hline
\end{tabular}
	\caption{Example of some IDataFrame functions supported by \ihpc{}.}
	\label{tab:functions}  
\end{table}

\section{Programming applications for \ihpc{}}
\label{sec:programming}

\ihpc{} requires a minimal driver code that implements the application at high-level. To facilitate the adoption from the Big Data community, the \ihpc{} API is inspired by the Spark API in such a way that \ihpc{} codes are easily understandable by users who are familiar with Spark. In comparison to Ignis, we have extended the API to cover the most important primitives required by Big Data applications. For instance, \ihpc{} includes functions such as \texttt{join} and \texttt{union} for graph processing. The \ihpc{} driver API is composed by six main classes:

\begin{itemize}[noitemsep]
    \item \texttt{Ignis} starts and stops the driver environment. 
    
    \item \texttt{IProperties} defines the execution environment properties. 
    
    \item \texttt{ICluster} represents a group of executors containers. It is possible, for example, to execute remote commands (\texttt{execute}, \texttt{executeScript}) and send files (\texttt{sendFile}, \texttt{sendCompressedFile}) to them.
    
   \item \texttt{IWorker} represents a group of processes of the same programming language. This class includes functions to read files (\texttt{textFile},  \texttt{partitionJsonFile}, etc.), import data partitions from another worker (\texttt{importData}), send data from the driver (\texttt{parallelize}) and execute external codes (\texttt{loadLibrary}, \texttt{call}, \texttt{voidCall}). As we explain later, the former routines allow \ihpc{} to execute MPI applications within the framework. 
   
    \item \texttt{IDataFrame} contains all the functions of the MapReduce paradigm, similarly to Spark RDD. A function can be a transformation that generates another IDataFrame or an action that generates a final result (see Table \ref{tab:functions}). With respect to Ignis, besides the support for new API functions, \ihpc{} has increased the overall performance for some types of routines (e.g., \emph{Group} and \emph{Sort}) thanks to its complete redesign using MPI. 
    
    \item \texttt{ISource} is an auxiliary class used by meta-functions such as \texttt{map} in the driver. This class acts as a wrapper for the input parameters. It is also used to store variables and send them to the executors. Those variables can be obtained by the executors using the context.
\end{itemize}

\subsection{An example: Transitive Closure}
\label{sec:example}

\begin{figure}[t!]
\begin{lstlisting}[language=Python]
 #!/usr/bin/python

 import ignis

 # Initialization of the framework
 ignis.Ignis.start()
 # Resources/Configuration of the cluster
 prop = ignis.IProperties()
 prop["ignis.executor.image"] = "ignishpc/full"
 prop["ignis.executor.instances"] = "2"
 prop["ignis.executor.cores"] = "4"
 prop["ignis.executor.memory"] = "2GB"
 # Construction of the cluster
 cluster = ignis.ICluster(prop)
 # Initialization of a Python Worker
 worker_python = ignis.IWorker(cluster, "python")
 # Task 1: Tokenize text into pairs (x, y)
 # The edges are stored in reversed order
 tc =  worker_python.textFile("graph.dat")
 edges = tc.map(lambda x_y: x_y.split(" ")[::-1])
 # Initialization of a C++  Worker
 worker_cpp = ignis.IWorker(cluster, "cpp")
 # Transfer data from Task 2 - Python
 tc2 = worker_cpp.importData(tc).cache()

 oldCount = 0
 nextCount = tc.count()
 while True:
   oldCount = nextCount
   # Task 2:
   # Perform the join,(y, (z, x)) pairs,
   # to obtain the new (x, z) paths.
   new_edges = tc2.join(edges).
      map("libexample.so:Reverse2")
   tc2 = tc2.union(new_edges).distinct().cache()
   nextCount = tc2.count()
   if nextCount == oldCount:
     break
 # Show result
 print("TC has %i edges" % tc2.count())
 # Stop the framework
 ignis.Ignis.stop()

\end{lstlisting}
	\vspace{-0.25cm}
	\caption{\emph{Transitive Closure} driver code in Python.}
	\label{fig:driver1}    
\end{figure}

With the goal of illustrating how applications are programmed in \ihpc{}, Figure \ref{fig:driver1} shows an example of a driver implemented in Python for computing the \emph{Transitive Closure} of a graph. This algorithm finds out if a vertex $x$ is reachable from another vertex $y$ for all vertex pairs $(x, y)$ in the graph. Note that an equivalent driver code could be implemented in any of the \ihpc{} supported languages (C/C++ and Java) using a similar syntax.

First, the \ihpc{} framework is initialized (line 6). Properties are created to configure and build a cluster (lines 8 to 14). Note that the properties definition is optional, and \ihpc{} could read them from a default configuration file. Moreover, \ihpc{} introduce the possibility of overwrite the default values when a job is submitted like Spark using the new \textsc{Submitter} module. Therefore, the Docker image, the number of containers, the number of cores and the memory per container could be defined out of the driver code. Computing the Transitive Closure has two phases. To illustrate the multi-language support in \ihpc{}, each phase was implemented in a different programming language. The first one uses a Python executor and the second a C++ executor. The first stage consists of a \texttt{map} operation that takes as input a text file and creates pair values that represent edges in the graph (line 20). As a consequence, it is necessary to previously create a Python worker in the cluster (line 16). It is important to note that creating the worker is mandatory and is not related to the driver programming language. On the other hand, if the worker and the driver code are in the same language, lambda functions can be used (line 20). 

The following phase of the algorithm is iterative: edges are joined into a path until there are not new paths. Since this phase is implemented in C++, a C++ worker should be created (line 22). Data is shared between workers using the \texttt{importData} function (line 24). The driver code ends printing the results. The framework must be stopped before the driver ends (line 42) to stop the backend. Unlike Ignis, \ihpc{} automatically detects when the driver process ends and stops it. However, this is not a good practice.

In \ihpc{} a lazy evaluation is performed when a result is no required explicitly. In the example, the trigger that causes the tasks to be launched are the calls to the \texttt{count} function. 
This approach is also followed by Spark where RDDs are computed lazily the first time they are used in an action~\cite{Zah12}.

\begin{figure}[t!]
\begin{lstlisting}[language=C++]
 #include<ignis/executor/api/function/IFunction.h>

 using namespace ignis::executor::api;
 typedef std::pair<int64_t, int64_t> ipair;
 typedef std::pair<int64_t, ipair> ipair2;

 // Reverse pairs (z, (x, y)) to (z, (y, x))
 class Reverse2 : public function::IFunction<
       ipair2, ipair2>{
   ipair2 call(ipair2& x_y, IContext& context){
      return {x_y.second, x_y.first};
 };

 ignis_export(Reverse2, Reverse2)
\end{lstlisting}
	\vspace{-0.25cm}
	\caption{Function in C++ used by a \texttt{map} operation for the \emph{Transitive Closure} application.}
	\label{fig:executor}    
\end{figure}

Most of the driver functions are meta-functions. That is, generic functions that require another one to perform an internal operation. This is the case of \texttt{map} in the example of Figure \ref{fig:driver1} (lines 20 and 33-34). To implement those functions we should use the executor API provided by \ihpc{}. Basically, it defines a simple interface based on the number of required input parameters. Figure \ref{fig:executor} shows an example corresponding to the C++ function used by \texttt{map} in the driver code. Since \texttt{map} takes one parameter and also returns one parameter, \texttt{IFunction} is used. In case there are two input parameters (e.g., \texttt{reduce}), \texttt{IFunction2} would be used, and so on. Note that if the function does not return any value (e.g. \texttt{foreach}), functions have the same name but with the \texttt{Void} prefix.

\subsection{Text lambda functions}

As explained previously, lambda functions need that driver and executor codes were implemented in the same language because native code serialization is required. Note that although Python and Java are able to serialize data both languages face compatibility problems. On top of that, C++ does not allow any type of code serialization. 
To overcome these limitations \ihpc{} implements its own multi-language lambdas without source code serialization, named \emph{text lambdas}. In this way, \ihpc{} allows to define lambdas as text, using the executor language syntax. The executor will transform the lambda text into source code to be used as a meta-function parameter. It is important to highlight that the language of the driver code is indifferent. 

\begin{figure}[t!]
\begin{lstlisting}[language=C]
 // Python lambda
 data.reduce("lambda a, b: a + b")

 // C++ lambda
 v = 2
 data.reduce(
   ISource("[](int a, int b, IContext &c){           \
             return (a + b) * c.var<int>(\"v\"); \
         }").addParam("v",v)
 )
\end{lstlisting}
	\vspace{-0.25cm}
	\caption{Examples of text lambda for a Python and a C++ executor.}
	\label{fig:multi_lambda}    
\end{figure}

Figure \ref{fig:multi_lambda} shows an example of a text lambda that accumulates all elements (line 2) used by a \texttt{reduce} function. It uses Python syntax so must be evaluated by a Python executor. Another example is shown in line 7. It defines a text lambda that captures the value of a variable, which is read from the context. This lambda function will be compiled and loaded by a C ++ executor. Performance is not affected by using text lambda functions but it can add some overhead to the compilation process, especially for C++.

In the same way, using the mechanism that allows to execute text lambda functions, \ihpc{} can send a complete job or application to the executors. This is possible thanks to \texttt{loadLibrary}, which can be used to execute a full source code as a \ihpc{} library. It has again a small impact on the compilation time. More details about \texttt{loadLibrary} are provided in Section \ref{sec:mpi_apps_ihpc} using MPI applications as use case.

\section{MPI on \ihpc{}}
\label{sec:integrating_mpi}

Our first prototype, Ignis, was limited to perform inter-process communications using only TCP sockets (different computing nodes) or shared memory (same node). However, \ihpc{} was completely redesigned to use MPI as backbone technology. As a consequence, all communications are internally implemented by MPI routines. As we explained previously in the paper, this change makes it possible for \ihpc{} to support more communication models and network architectures. In addition, an important advantage of our approach is that, once \ihpc{} has configured the MPI communications, users can combine in the same application pure MPI libraries using the \ihpc{} communicators together with standard Big Data functions (\texttt{map}, \texttt{reduce}, \texttt{collect}, etc.).

\subsection{Integration of MPI into a Big Data environment}

MPI was not designed to run on Docker containers. As a consequence, there are several problems that should be addressed. First, by default and to preserve an isolation runtime environment, Docker creates a private virtual network between the host and the containers. If two containers are launched on the same host, we can execute MPI processes in the same way that they were two real nodes of a cluster. But if we launched those containers on different hosts, the communication is impossible since they belong to different networks. We can find in the literature several works that deal with this issue (see the Related Work section). For instance, some approaches opt for launching the container on the host network or creating a virtual network between the cluster nodes~\cite{Bay17}. 
However, these configurations are difficult to handle and implement by resource managers in Big Data environments.

Second, there are important differences in how ports are handled by a Big Data or a HPC environment. For instance, ports are considered a resource in a Big Data environment because there are services that require an exclusive port, which is not the case in HPC. MPI needs ports to establish connections between processes but restricted to a range. However, ports provided by resource managers in a Big Data environment are usually random and not consecutive. 

Finally, \ihpc{} can internally spawn several threads per MPI process (executor) to increase the performance when processing and/or communicating data. All these threads use communicators to exchange data in parallel. Every time a communicator is created, MPI assigns a virtual interface to it. However, a virtual interface can only be used by one communicator at the same time, so parallel communications requires the use of multiple virtual interfaces. In the most recent MPICH version, which is the MPI implementation used by \ihpc{}, virtual interfaces are assigned sequentially. Since \ihpc{} creates and destroys communicators dynamically, it is not possible to assure that threads can always exchange data in parallel using communicators with different assigned virtual interfaces. The consequence is a degradation in the performance.

To overcome the above problems, \ihpc{} applies the following changes to MPICH:

\begin{itemize}[noitemsep]
    \item Containers: MPICH has been designed to work on a local network. Docker containers can be joined to a network but only within the same node (internal network). Although resource managers can export a service from the internal network to the local network, this causes a problem when MPI is executed containerized. MPICH uses a service to store the network addresses of the launched MPI processes, but when using containers, each MPI process stores the values corresponding to the internal network which are not valid outside the node. For this reason, it is necessary to modify MPICH in order to store the correct network values that correspond to the local network.  
    
    \item Ports: now MPICH uses a list of ports provided by the resource manager instead of a range.
    
    \item Multithreading: MPICH was modified to assure that all threads use a different virtual connection. In this way, communications between threads can always be performed in parallel.
\end{itemize}

\subsection{Running MPI applications in \ihpc{}}
\label{sec:mpi_apps_ihpc}

\ihpc{} can execute MPI applications implemented in any of the supported languages. We must highlight that, to the best of our knowledge, currently does not exist other Big Data framework with this feature. Most of the MPI codes for HPC are implemented in C/C++, while applications in other languages such as Java and Python are a minority. So although \ihpc{} supports Python and Java, we will focus on C/C++ MPI applications. 

To explain how to execute a MPI application within \ihpc{}, we have considered LULESH~\cite{Kar13} as guiding example. LULESH is a proxy HPC application for shock hydrodynamics with more than 5,000 lines of C++ code. 

MPI applications, like other \ihpc{} codes, must be implemented using the executor API to be used from the driver (see Figure \ref{fig:executor}). However, some minimal modifications should be previously applied to the original MPI codes: 

\begin{figure}[t!]
\begin{lstlisting}[language=C]
 // mpi.h
 #ifndef IGNIS_MPI
 #define IGNIS_MPI

 #include_next <mpi.h>

 extern MPI_Comm IGNIS_COMM_WORLD;
 #undef MPI_COMM_WORLD
 #define MPI_COMM_WORLD IGNIS_COMM_WORLD

 #endif
\end{lstlisting}
	\vspace{-0.25cm}
	\caption{C header that replaces \texttt{MPI\_COMM\_WORLD} by \texttt{IGNIS\_COMM\_WORLD} (global variable).}
	\label{fig:mpi_header}    
\end{figure}

\begin{itemize}[noitemsep]
    \item MPI initialization: \ihpc{} controls the MPI environment, so \texttt{MPI\_Init} and \texttt{MPI\_Finalize} must be removed from the MPI application.
    \item \texttt{MPI\_COMM\_WORLD}: MPI applications use a default communicator but \ihpc{} requires its own communicator. The simplest solution is to create a custom header to overwrite the default communicator. Figure \ref{fig:mpi_header} shows an implementation of this functionality.
    \item I/O data: These modifications are optional. In some scenarios, it is interesting to allow \ihpc{} to handle the operations on input and output data of an MPI application. This is the case, for example, when the output file of the application will be afterwards processed by other \ihpc{} tasks. If \ihpc{} manages the output file, data is kept in memory. If not, the output file would be written to disk and read again to continue executing the following tasks, causing an important degradation in the performance. To do that, read and write functions related to input/output files will be removed from the MPI source code. As we explain next, they will be replaced by input and return parameters of the \texttt{call} function in the corresponding executor code. 
\end{itemize}

\begin{figure}[t!]
\begin{lstlisting}[language=C++]
 using ignis::executor::api;

 extern int main(int argc, char *argv[]);
 MPI_Comm IGNIS_COMM_WORLD;
 class Lulesh : public function::IVoidFunction0{
 public:
   void call(IContext &context) override {

    IGNIS_COMM_WORLD = context.mpiGroup();
    const char *argv[100];
    argv[0] = "ignis-cpp";
    int argc = 1;
  
    // Start parsing arguments ****************
    if (context.isVar("i")) {
     argv[argc++] = "-i";
     argv[argc++] = const_cast<char *>(
         context.var<std::string>("i").data());
    }

    if (context.isVar("s")) {
     argv[argc++] = "-s";
     argv[argc++] = const_cast<char *>(
         context.var<std::string>("s").data());
    }
  
    ...
  
    if (context.isVar("h") && context.var<bool>("h")) {
     argv[argc++] = "-h";
    }
   // End parsing arguments ****************
  
   main(argc, const_cast<char **>(argv)); 
   }
  };

 ignis_export(lulesh_app, Lulesh)

 create_ignis_library("lulesh_app");
\end{lstlisting}
	\vspace{-0.25cm}
	\caption{Executor code for LULESH using C++.}
	\label{fig:lulesh_call}    
\end{figure}

Figure \ref{fig:lulesh_call} shows the executor code for calling LULESH from the \ihpc{} driver. First, the global variable for the communicator of Figure \ref{fig:mpi_header} is created (line 4) and initialized with the \ihpc{} MPI group (line 9). LULESH is a benchmark, so it does not receive any data from \ihpc{}. As a consequence, according to the executor API explained in Section \ref{sec:programming}, it is of type \texttt{IVoidFunction0} (line 5). The only mandatory method to be implemented is \texttt{call} (line 7). Inside that method, the application arguments are parsed from the \ihpc{} context. In our example, each argument is individually parsed to create a user friendly interface. However, the arguments could be parsed together as a list, reducing noticeably the necessary lines of code. The \texttt{call} method ends calling the LULESH main function (line 34). Afterwards, \texttt{Lulesh} class is exported (line 38). This operation is only necessary in C++. To finalize, an \ihpc{} library is created (line 40), which will be used to call LULESH from the driver.

\begin{figure}[t]
\begin{lstlisting}[language=Python]
 worker.loadLibrary("liblulesh.so")
 worker.voidCall("lulesh_app",  s = "70", i = "2400")


\end{lstlisting}
\begin{lstlisting}[language=C++]
 worker.loadLibrary("liblulesh.so");
 worker.voidCall(ISource("lulesh_app").
                    addParam("s","70").
                    addParam("i","2400"));
\end{lstlisting}
	\vspace{-0.25cm}
	\caption{Lulesh usage from a Python driver (top) and its equivalent C++ code (bottom).}
	\label{fig:lulesh_driver}    
\end{figure}

Finally, we show how to use an MPI application from the driver. The example of Figure \ref{fig:lulesh_driver} focuses only in the necessary functions to execute LULESH using a Python and a C++ driver. Note that the MPI application should be previously compiled as a library (\texttt{liblulesh.so}). The example assumes that there is a C++ worker in the driver where two functions are executed: \texttt{loadLibrary} and \texttt{voidCall}. On the one hand, \texttt{loadLibrary} loads all the classes from the library declared in \texttt{create\_ignis\_library}. In this case, \texttt{Lulesh} is the only existent class. On the other hand, \texttt{voidCall} is an action that causes the execution of the library. If the library returns an output to \ihpc{}, \texttt{voidCall} should be replaced by \texttt{call} that would return a \texttt{IDataFrame} object. Library arguments in C++, which were parsed in the executor code, are added to the function using \texttt{addParam} from the \texttt{ISource} class. This syntax could be also used in Python. Nevertheless, keyword arguments in Python are a more elegant alternative (see line 2 in Figure \ref{fig:lulesh_driver}).

\subsection{Hybrid applications}

\begin{figure}[t]
\begin{lstlisting}[language=Python]
 ...
 # Initialization of a Python Worker
 worker = ignis.IWorker(cluster, "python")
 # Load wordcount function implemented in Python with MPI
 worker.loadLibrary("wordcount.py")
 # Task 1: Read a text line by line
 text = worker.textFile("text.txt")
 # Task 2: Split each line into words
 words = text.flatmap(lambda line : line.split(" "))
 # Task 3: Execute wordcount using words as input
 counts = worker.call("wordcount", words)
 # Task 4: Result is stored as json
 counts.saveAsJson("result.json")
 ....
\end{lstlisting}
	\vspace{-0.25cm}
	\caption{Wordcount example as MPI hybrid application.}
	\label{fig:mpi_hybrid}    
\end{figure}

In \ihpc{} an MPI code can be combined with typical MapReduce operations to create a hybrid application. In this way, the different computing tasks could be implemented in the programming model and language that best suits them. 

Figure \ref{fig:mpi_hybrid} shows a simple example of a Wordcount application where a MPI Python library is combined with \ihpc{} API functions. Input data is distributed and prepared by \ihpc{} (Tasks 1 and 2), so MPI is only responsible of the compute-intensive part (Task 3). Observe that for using functions included in a Python library it is only necessary to load the library (line 5) and invoke the \texttt{call} routine with the name of the desired function (line 11). Finally, results are converted and written to disk in \emph{json} format using the \ihpc{} API (Task 4).

\section{Experimental evaluation}
\label{sec:exp_eval}

\subsection{Experimental setup}

The experiments shown in this section were carried out on an 12-node cluster, where each node consists of:
\begin{itemize}[leftmargin=*,noitemsep]
\item CPU: 2 $\times$ Intel Xeon E5-2630v4 (2.2Ghz, 10 cores)
\item Memory: 384 GB of RAM
\item Storage: 8 $\times$ 4TB 7.2k SATA
\item Network: 2 $\times$ 10GbE
\end{itemize}

It is a Linux cluster running CentOS 7 (kernel 3.10.0), Docker 20.10.2-ce and Spark 2.2.0 (with YARN~\cite{Vav13} as cluster manager). Ignis and \ihpc{} run on a Ubuntu 20.04 image with MPICH 3.4.1.

\begin{table}[t!]
\footnotesize
\centering
\begin{tabular}{lC{0.8cm}C{0.8cm}C{0.8cm}C{0.8cm}C{0.8cm}}
\hline
\rowcolor{lightgray}
& \multicolumn{2}{c}{\bf Batch (one pass)} & \multicolumn{3}{c}{\bf Iterative (caching)}  \\ \cline{2-6} 
\rowcolor{lightgray}
\multirow{-2}{*}{\bf Operators} & {\bf MB}  & {\bf TS}  & {\bf KM}  &  {\bf PR}  &   {\bf TC} \\ \hline
\texttt{textFile}        & \checkmark   & \checkmark  & \checkmark & \checkmark & \checkmark \\ \hline
\texttt{map}             & \checkmark   & \checkmark  & \checkmark & \checkmark & \checkmark \\ \hline
\texttt{mapValues}       &      --      &      --     &     --     & \checkmark &     --     \\ \hline
\texttt{flatMap}         &      --      &      --     &     --     & \checkmark &     --     \\ \hline
\texttt{reduceByKey}     &      --      &      --     & \checkmark & \checkmark &     --     \\ \hline
\texttt{collectAsMap}    &      --      &      --     & \checkmark &     --     &     --     \\ \hline
\texttt{repartition}     &      --      & \checkmark  &     --     & \checkmark &     --     \\ \hline
\texttt{count}           &      --      &      --     &     --     & \checkmark & \checkmark \\ \hline
\texttt{join}            &      --      &      --     &     --     & \checkmark & \checkmark \\ \hline
\texttt{union}           &      --      &      --     &     --     &     --     & \checkmark \\ \hline
\texttt{distinct}        &      --      &      --     &     --     &     --     & \checkmark \\ \hline
\texttt{importData} (I)  &  \checkmark  &      --     &     --     &     --     &     --     \\ \hline
\texttt{sort}            &      --      & \checkmark  &     --     &     --     &     --     \\ \hline
\texttt{saveAsTextFile}  & \checkmark   & \checkmark  & \checkmark &     --     &     --     \\ \hline
\end{tabular}
	\caption{Operations used in each Big Data application. Minebench (MB), Terasort (TS), K-means (KM), PageRank (PR) and Transitive Closure (TC). Operators annotated with I are specific only to \ihpc{}.}
	\label{tab:mapreduce_calls}  
\end{table}

\subsection{Big Data applications}

We have selected five 
workloads that represent different types of application patterns for which Spark is considered the best performing Big Data framework: \emph{Minebench}, \emph{TeraSort}, \emph{K-Means}, \emph{PageRank} and \emph{Transitive Closure}. 
Table \ref{tab:mapreduce_calls} lists the use of the most important operators by each Big Data application, including basic core operators and specific ones implemented by the \ihpc{} framework.

\begin{itemize}[leftmargin=*,noitemsep]

  \item \emph{Minebench (MB)}. This application\footnote{Publicly available at: \url{https://github.com/brunneis/minebench}} performs the calculation of SHA-256 hashes imitating the Proof-of-Work algorithm used in the Bitcoin protocol~\cite{Bitcoin2008}. Do not confuse with the data mining benchmark suite NU-MineBench. This algorithm has two phases which are implemented using two chained map operations. The first map is data-intensive, while the second is a compute-intensive task. In particular, in the first stage a set of Bitcoin transactions are grouped together forming a block proposal. A binary Merkle tree~\cite{Merkle} is calculated for those transactions and its Merkle root hash is added to a block header. The second stage calculates the hash of the block header iteratively while the condition is not met. The strong scaling tests were obtained using an input file containing 300K blocks (120MB), while the weak scaling experiments fixed the input data per core to 300K blocks. 

  \item \emph{TeraSort (TS)}. It is a sorting algorithm suitable for measuring the I/O and the communication performance of the considered frameworks. Elements in \ihpc{} are sorted by means of the MergeSort algorithm where elements are distributed by a regular sampling among the executors~\cite{Li1993}. Note that this task requires that executors exchange data. The input data used in the tests contains 1 TB of text with 1.8B lines. 
 
\begin{figure*}[t!]
	\centering
	\subfloat[Times (strong scaling)]{\includegraphics[width=0.32\textwidth]{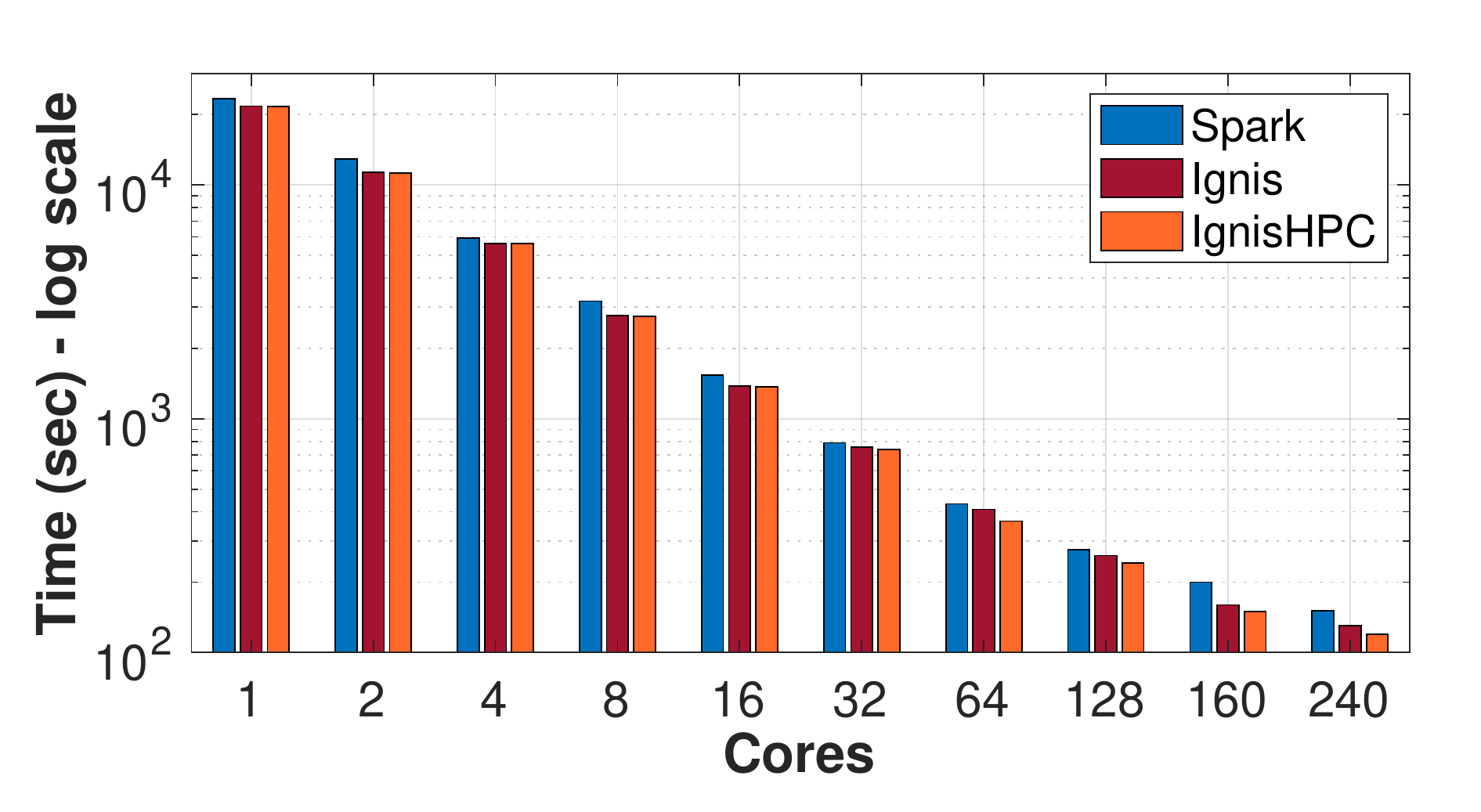}}
	\subfloat[Speedup]{\includegraphics[width=0.32\textwidth]{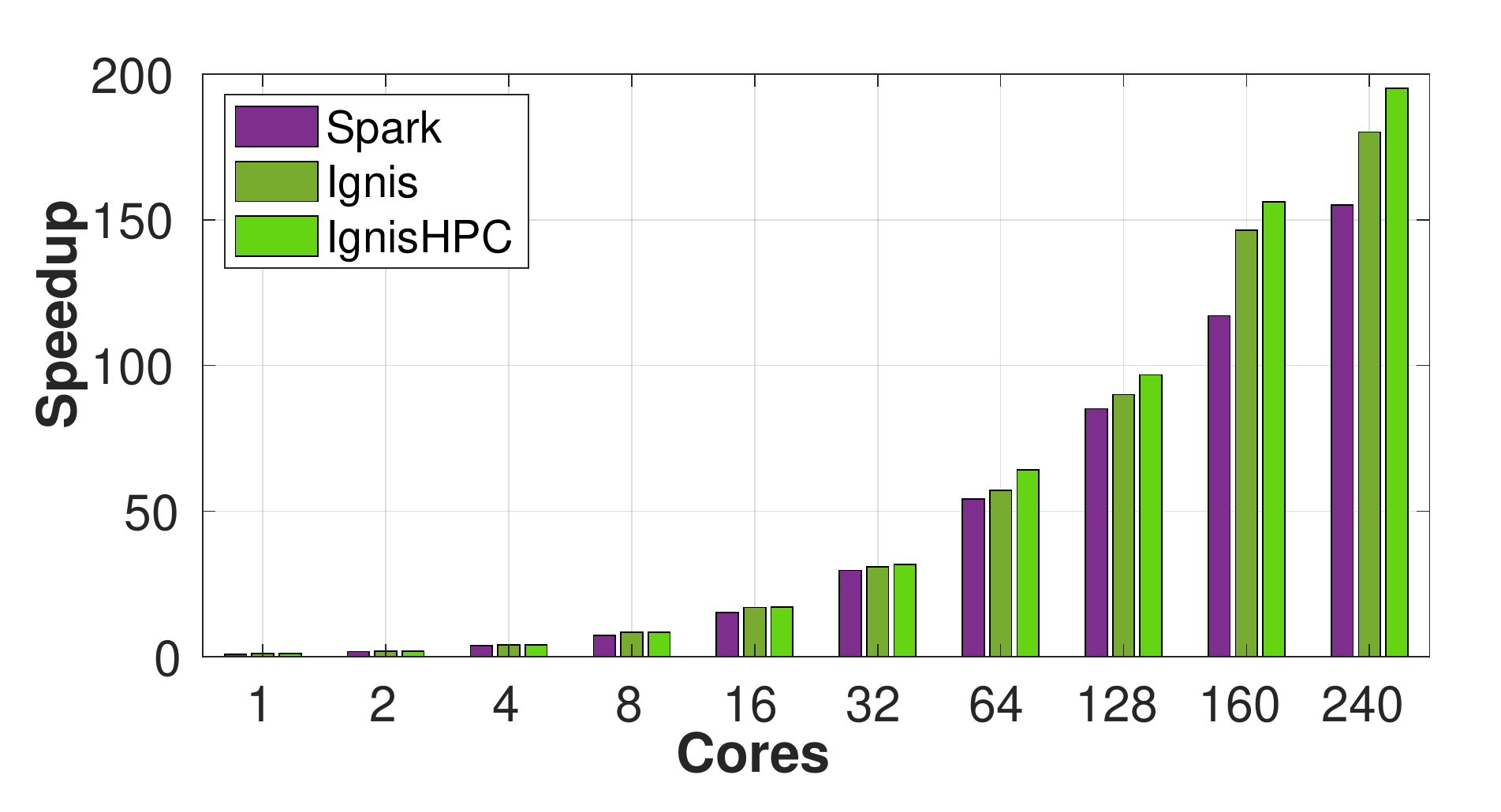}}
	\subfloat[Times (weak scaling)]{\includegraphics[width=0.32\textwidth]{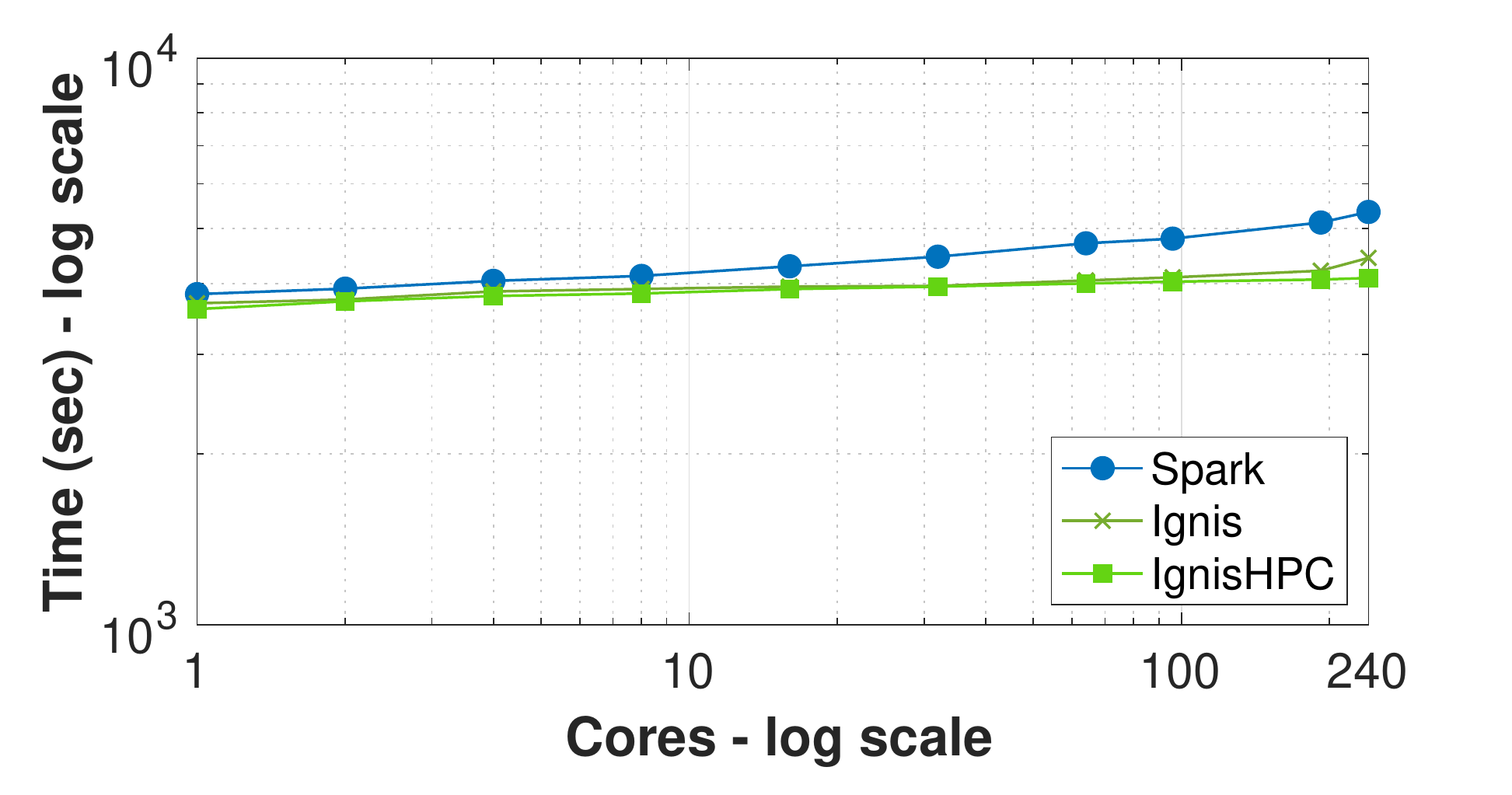}}
	\vspace{-0.25cm}
	\caption{Study of the scalability of \ihpc{}, Ignis and Apache Spark running the Python Minebench application.}
	\label{fig:py_minebench} 
\end{figure*}
\begin{figure*}[t!]
	\centering
	\subfloat[Times (strong scaling)]{\includegraphics[width=0.32\textwidth]{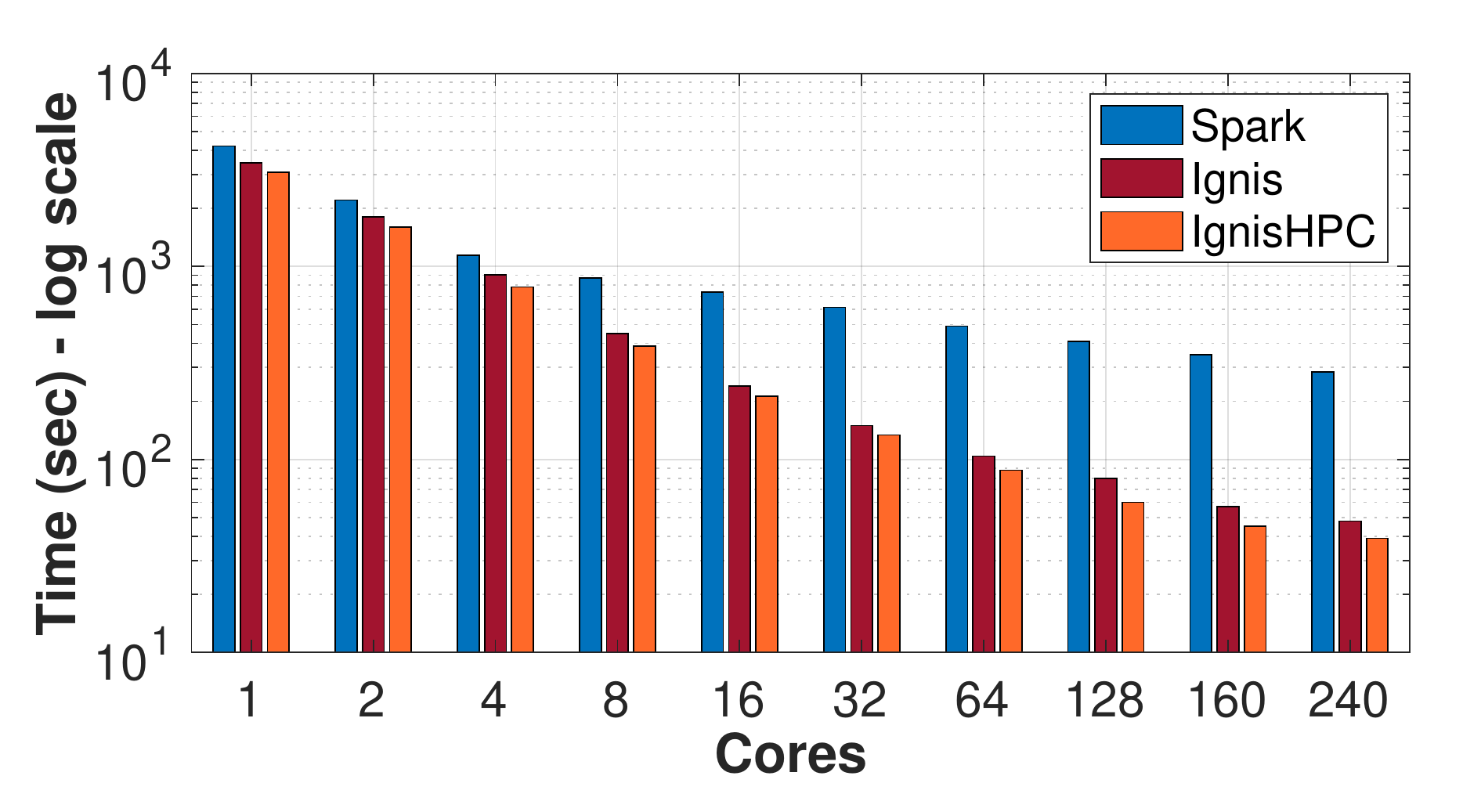}}
	\subfloat[Speedup]{\includegraphics[width=0.32\textwidth]{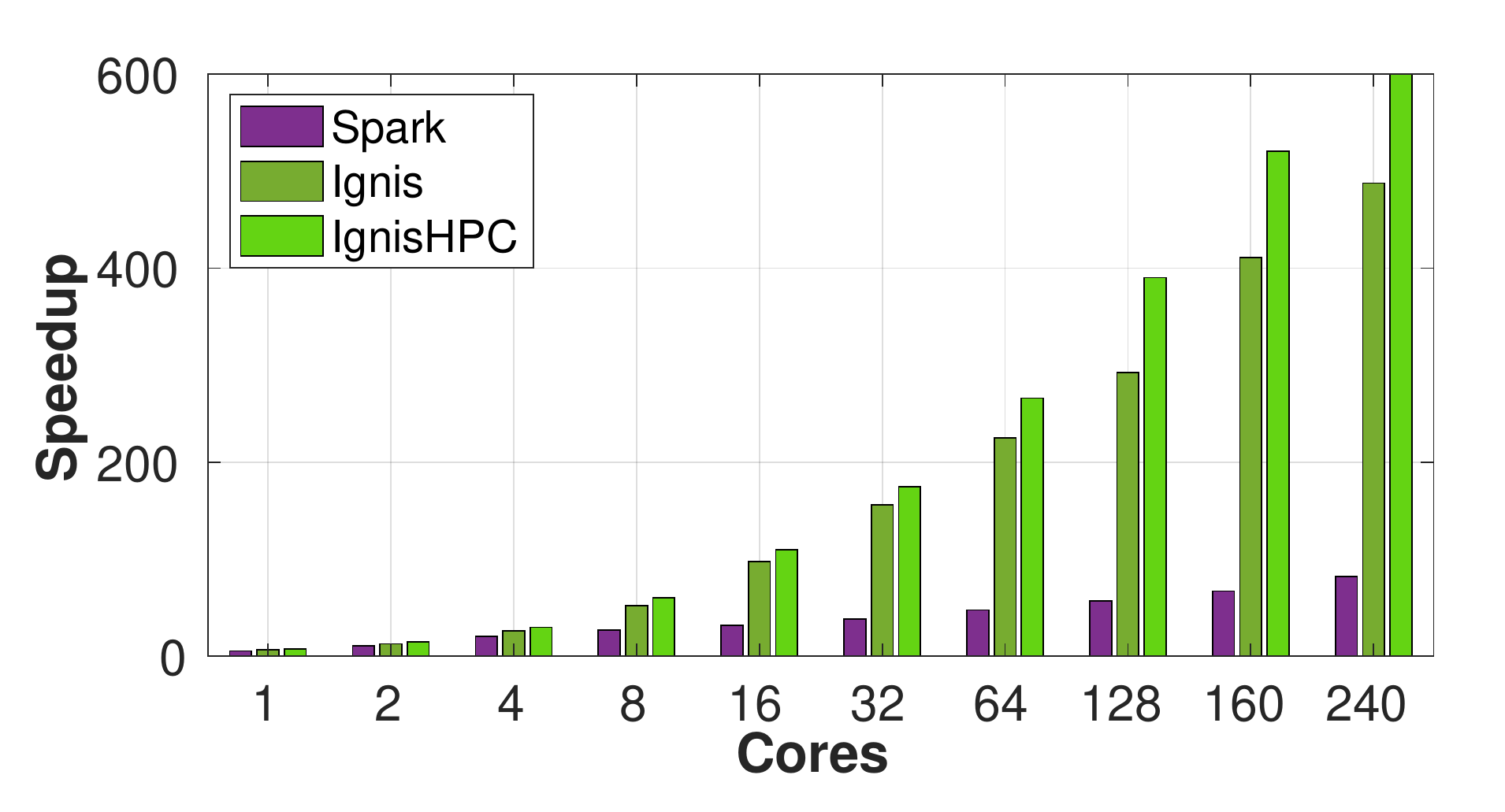}}
	\subfloat[Times (weak scaling)]{\includegraphics[width=0.32\textwidth]{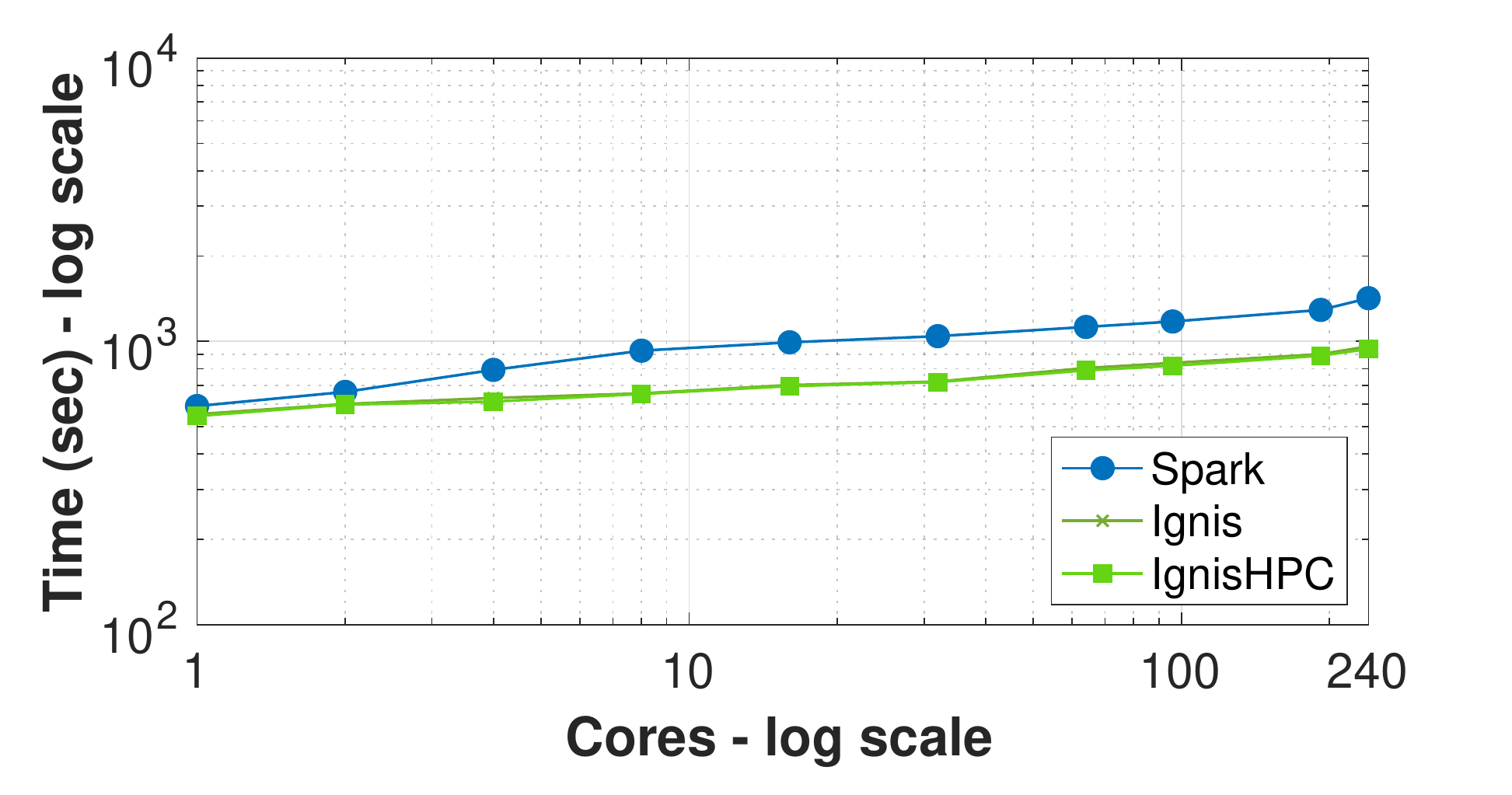}}
	\vspace{-0.25cm}
	\caption{Study of the scalability of \ihpc{}, Ignis and Apache Spark running the Python \& C++ Minebench application.}
	\label{fig:c++_minebench} 
\end{figure*}

  \item \emph{K-Means (KM)}. This is a classical machine learning algorithm for data clustering, and it is a good example of an iterative MapReduce application. This pattern covers a large set of iterative machine learning algorithms such as linear regression, logistic regression, and support vector machines. The goal of KM is to classify a given data set through a certain number of clusters ($K$ clusters). In each iteration, a data point is assigned to its nearest cluster center, using a map function. Data points are grouped to their center to further obtain a new cluster center at the end of each iteration (reduce). The experimental evaluation was carried out using the NUS-WIDE dataset~\cite{Chu09}, which contains 269,648 images with 500 attributes per image. In the tests the results were obtained after 10 iterations and $K=81$.

  \item \emph{PageRank (PR)}. It is a graph algorithm which ranks elements by counting the number and quality of links. To evaluate the PR algorithm on \ihpc{} and Spark we used the LiveJournal graph from the SNAP repository~\cite{snap}, which contains 4.8M vertices and about 69M edges.

  \item \emph{Transitive Closure (TC)}. One of the most basic questions that arises when analyzing a complex graph $G$ is whether one vertex $x$ can reach another vertex $y$ via a directed path. A way to store this information is to construct another graph, such that there is an edge $(x, y)$ in the new graph if and only if there is a path from $x$ to $y$ in the input graph. This new graph is called the Transitive Closure of $G$. Since computing the TC is very costly, we used a small graph with 75 vertices and 200 edges in our tests.
  
  
\end{itemize}

\begin{figure*}[t!]
	\centering
	\subfloat[Times (strong scaling)]{\includegraphics[width=0.32\textwidth]{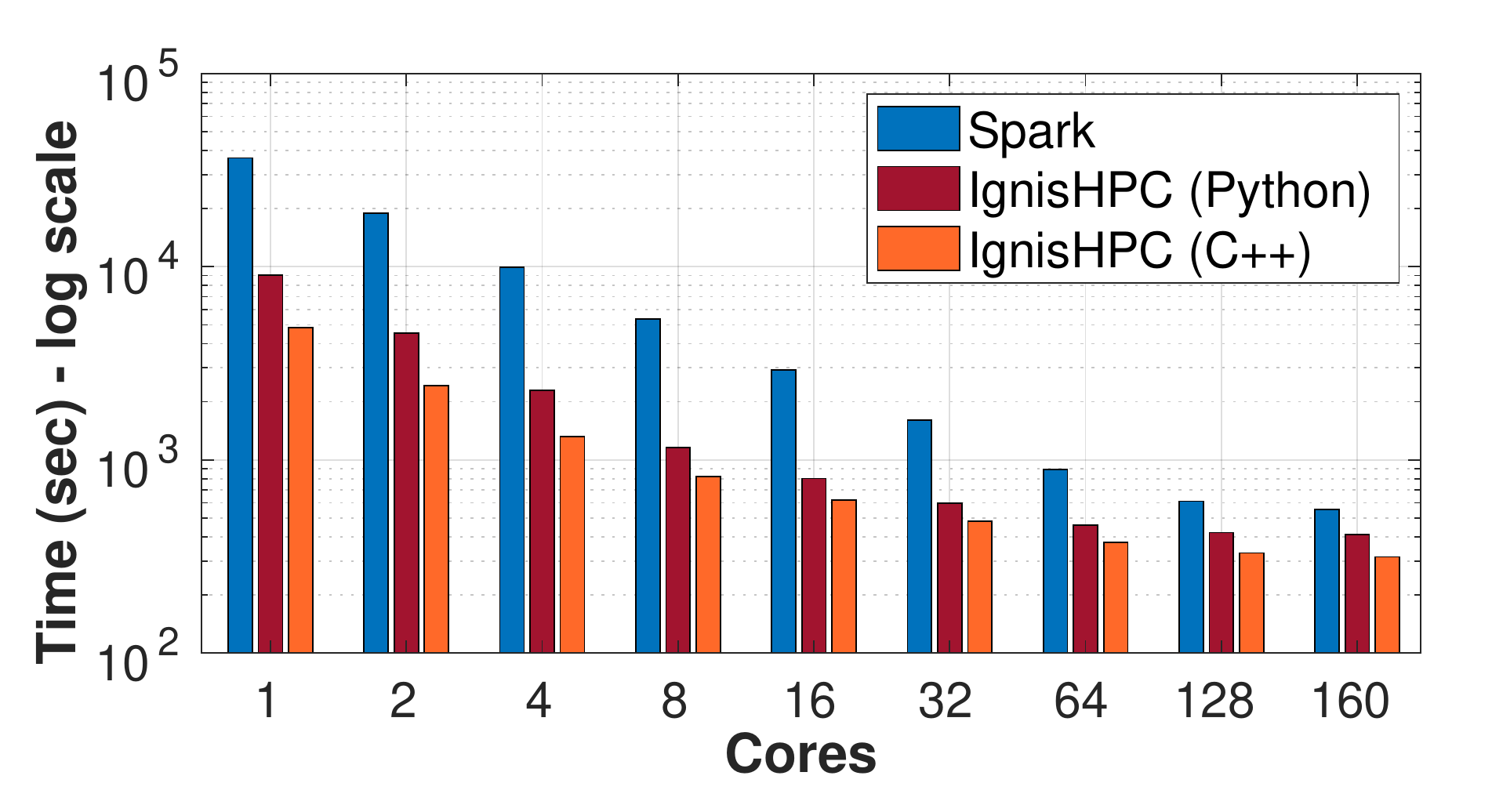}}
	\subfloat[Speedup]{\includegraphics[width=0.32\textwidth]{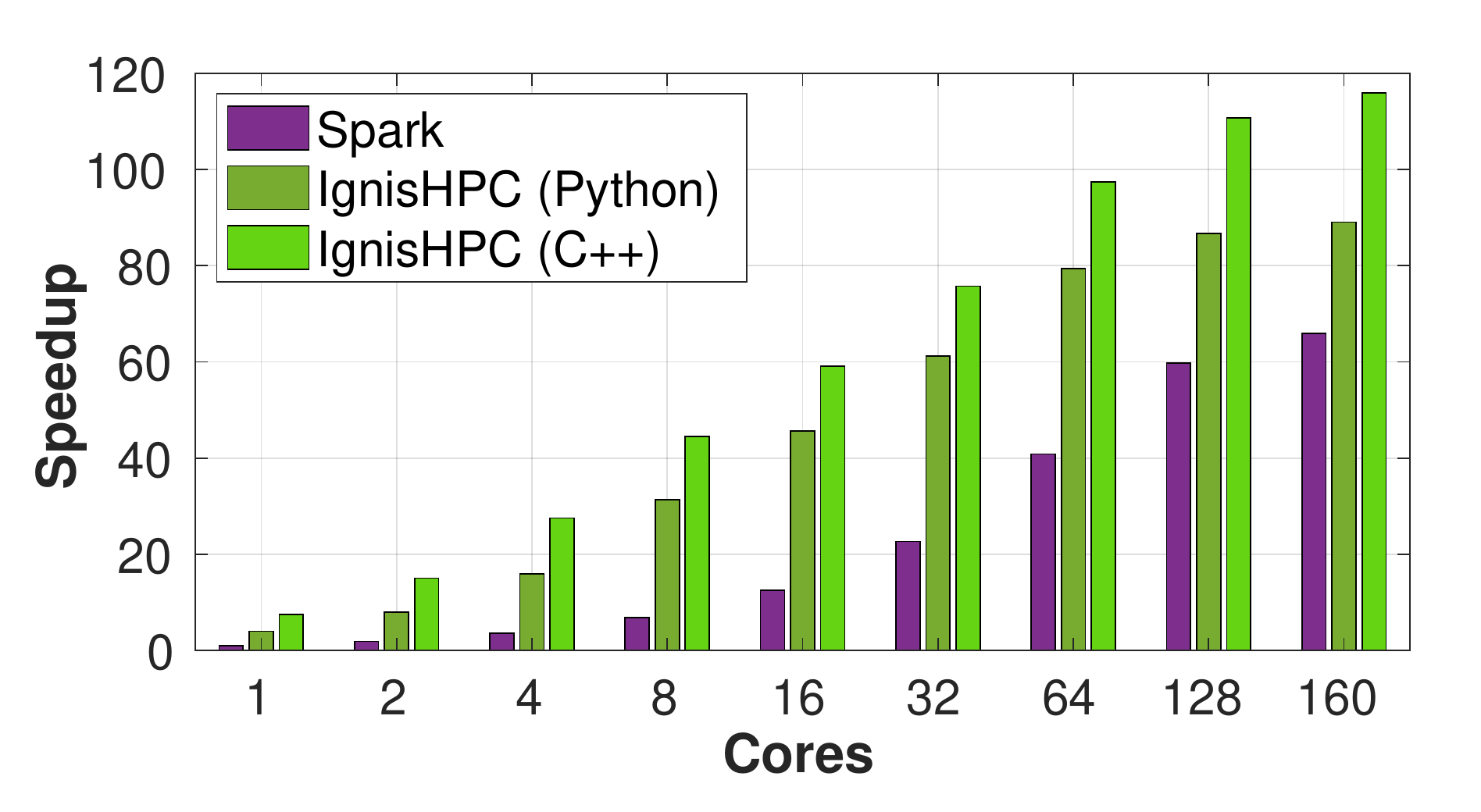}}
	\vspace{-0.25cm}
	\caption{Study of the scalability of \ihpc{} and Apache Spark running the TeraSort application.}
	\label{fig:terasort} 
\end{figure*}
\begin{figure*}[t!]
	\centering
	\subfloat[Times (strong scaling)]{\includegraphics[width=0.32\textwidth]{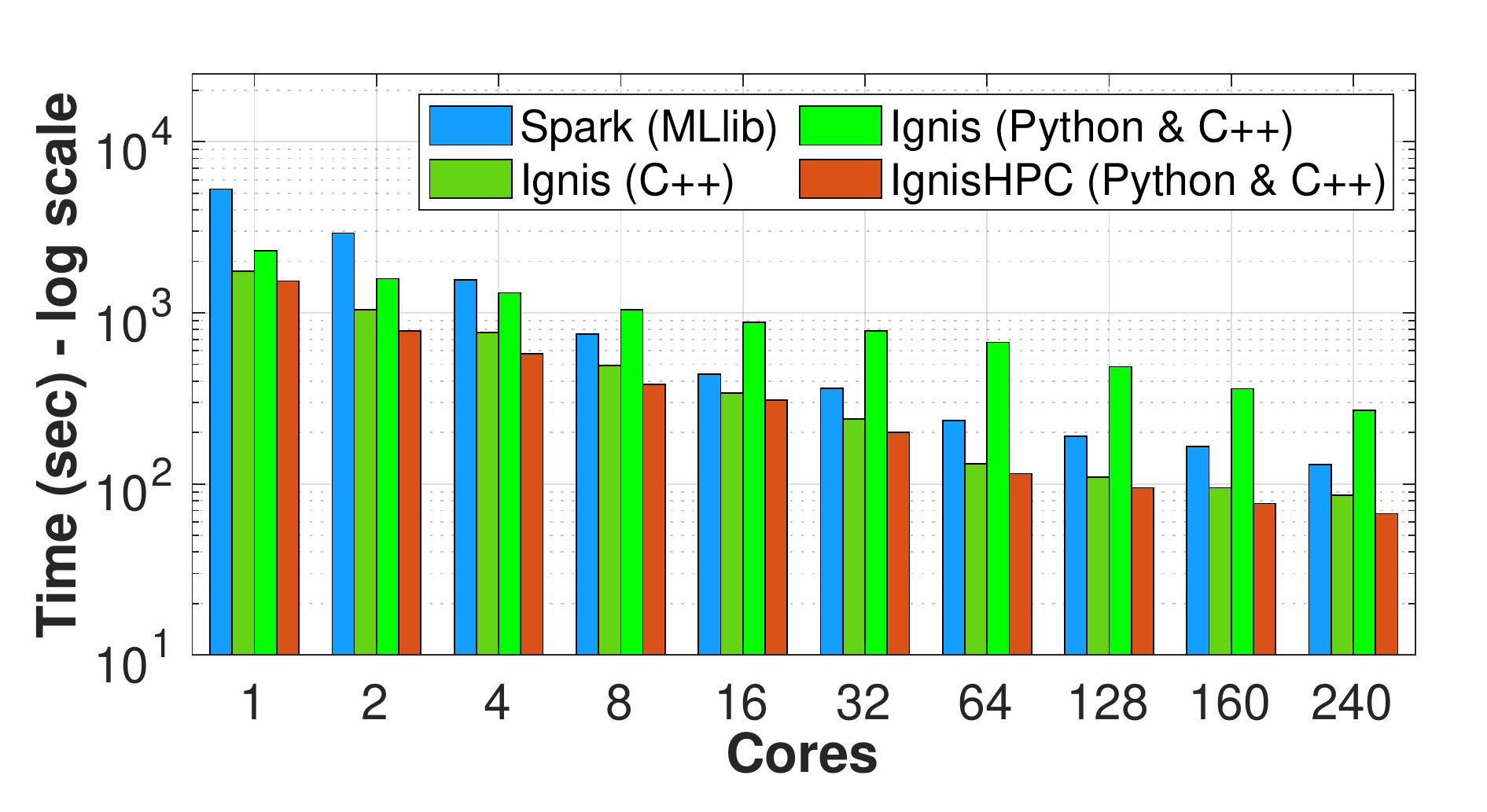}}
	\subfloat[Speedup]{\includegraphics[width=0.32\textwidth]{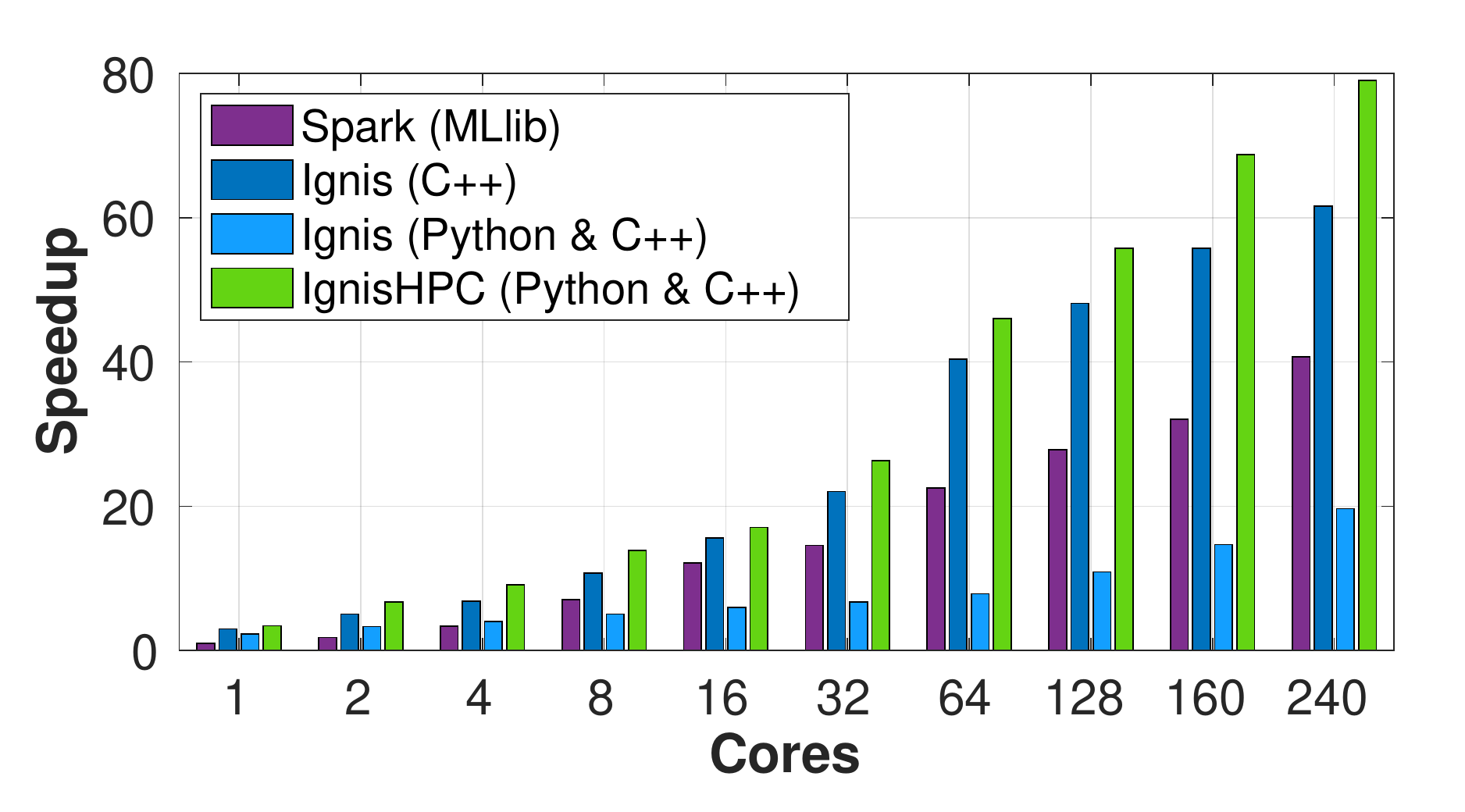}}
	\vspace{-0.25cm}
	\caption{Study of the scalability of \ihpc{}, Ignis and Apache Spark running the K-Means application.}
	\label{fig:kmeans} 
\end{figure*}

\subsubsection{Analysis and discussion}

We now present the performance results from our evaluation of Spark, Ignis and \ihpc{} considering all the Big Data applications detailed above. Figures \ref{fig:py_minebench} and \ref{fig:c++_minebench} show the scalability study of the \emph{Minebench} (MB) application using two different implementations. In the first one, MB was programmed using only Python, while in the second one, the two chained map operations are implemented using Python (data-intensive task) and C++ (compute-intensive task), respectively. Results show that \ihpc{} exhibits very good strong-scaling behavior for both implementations. On the contrary, the Spark scalability is impacted for the cost of starting JVMs and transferring data through system pipes to the Python processes, causing an important degradation in the overall performance~\cite{Pin20}. This scenario is even more clear for the multi-language implementation in Figure \ref{fig:c++_minebench}(a) since Spark sends data from Python to C++ processes through the JVM, increasing the number of pipe operations. As a consequence, the Spark strong scalability is really poor. On the other hand, \ihpc{} weak scales very well for both code versions (Figures \ref{fig:py_minebench}(c) and \ref{fig:c++_minebench}(c)), which is not surprising since there is not much communication in the MB application. As a consequence, \ihpc{} is able to extract all the existent parallelism. Finally, we must highlight that \ihpc{} clearly outperforms Ignis both in terms of strong and weak scalability, especially when considering the multi-language application.

Performance results of the TeraSort (TS) application running on \ihpc{} and Spark frameworks are displayed in Figure \ref{fig:terasort}. Ignis results are not shown because the memory consumption of sorting 1 TB of data is too high for our cluster. As we explained in Section \ref{sec:data_storage}, Ignis assigns one data partition to each executor. For TS those partitions are very large. Every time an element is added to a partition, the complete partition is copied to a different memory location (realloc operation), which causes a boost in the memory requirements. This restricts Ignis to work with smaller input datasets. We avoid this limitation since \ihpc{} was designed to allow several partitions per worker. Two different TS implementations in \ihpc{} were analyzed: a pure Python code and a multi-language Python-C++ code. In the latter case, the sort operation uses a user-defined C++ function for comparison purposes. For all the cases \ihpc{} outperforms Spark, especially when considering the multi-language implementation. In this way, for instance, \ihpc{} is 116$\times$ faster than sequential Spark when using 160 cores, while Spark reaches a speedup of only 66$\times$ (Figure \ref{fig:terasort}(b)). It allows \ihpc{} to sort 1 TB of data in barely 5 minutes.

\begin{figure*}[t!]
	\centering
	\subfloat[Times (strong scaling)]{\includegraphics[width=0.32\textwidth]{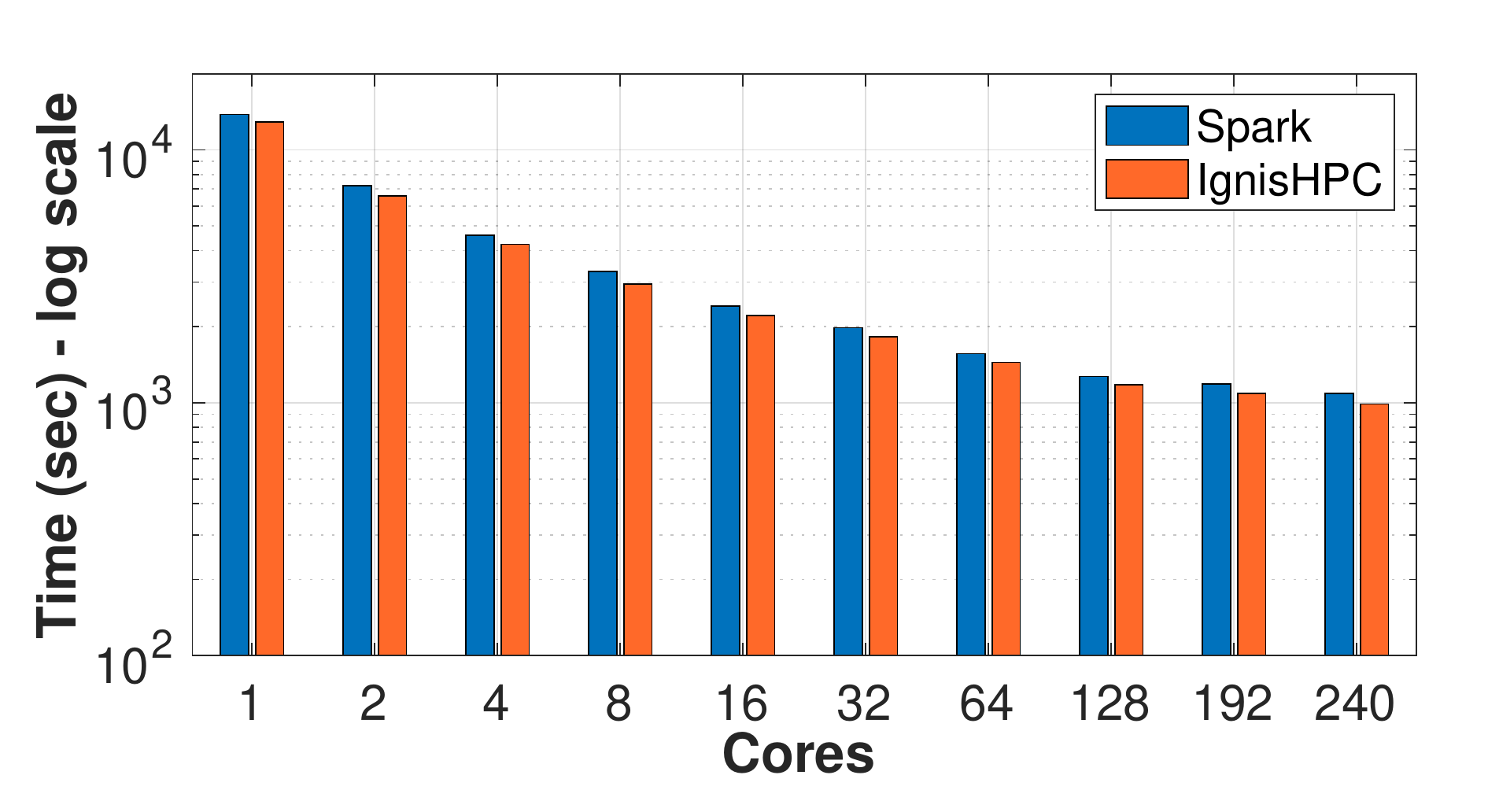}}
	\subfloat[Speedup]{\includegraphics[width=0.32\textwidth]{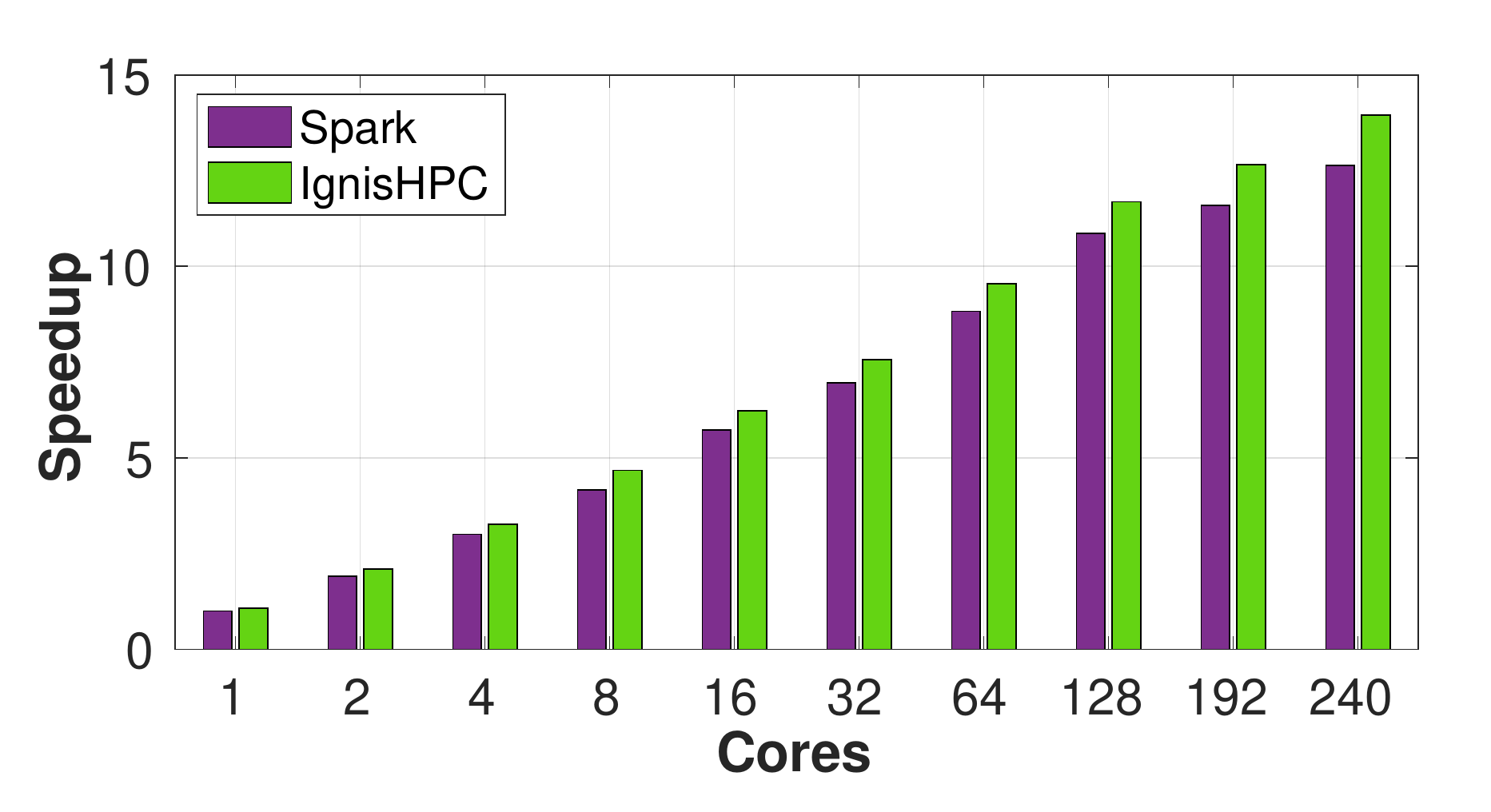}}
	\vspace{-0.25cm}
	\caption{Study of the scalability of \ihpc{} and Apache Spark running the PageRank application.}
	\label{fig:pagerank} 
\end{figure*}

Strong scaling results of K-Means (KM) are shown in Figure \ref{fig:kmeans}. We used as reference the Spark implementation of this algorithm included in MLlib (Machine Learning Library)~\cite{Men16}. For Ignis, a pure C++ and a Python-C++ implementations of KM were analyzed. For \ihpc{}, we only show the results for the multi-language Python-C++ code because the numbers obtained by a pure C++ application are very similar. We can observe that Ignis was able to beat Spark when considering the C++ code. However, an important degradation in the scalability was detected for the multi-language implementation as the parallelism increases. This problem was explained in Section \ref{sec:executor} and is related to the way Ignis handle iterative applications. Ignis starts and stops the executors each iteration because the driver must compute the partial results, which has an important impact on the performance. To deal with this, \ihpc{} takes advantage of MPI in such a way that executors compute the partial results and share them without intervention of the driver. For this reason \ihpc{} exhibits a very good strong scalability even for multi-language iterative applications, decreasing noticeably the execution times with respect to Spark and Ignis. It can be observed in Figure \ref{fig:kmeans}(b) that \ihpc{} is about two and four times faster than Spark and Ignis (multi-language code) when using all the cores in the cluster, respectively.  

\begin{figure*}[t!]
	\centering
	\subfloat[Times (strong scaling)]{\includegraphics[width=0.32\textwidth]{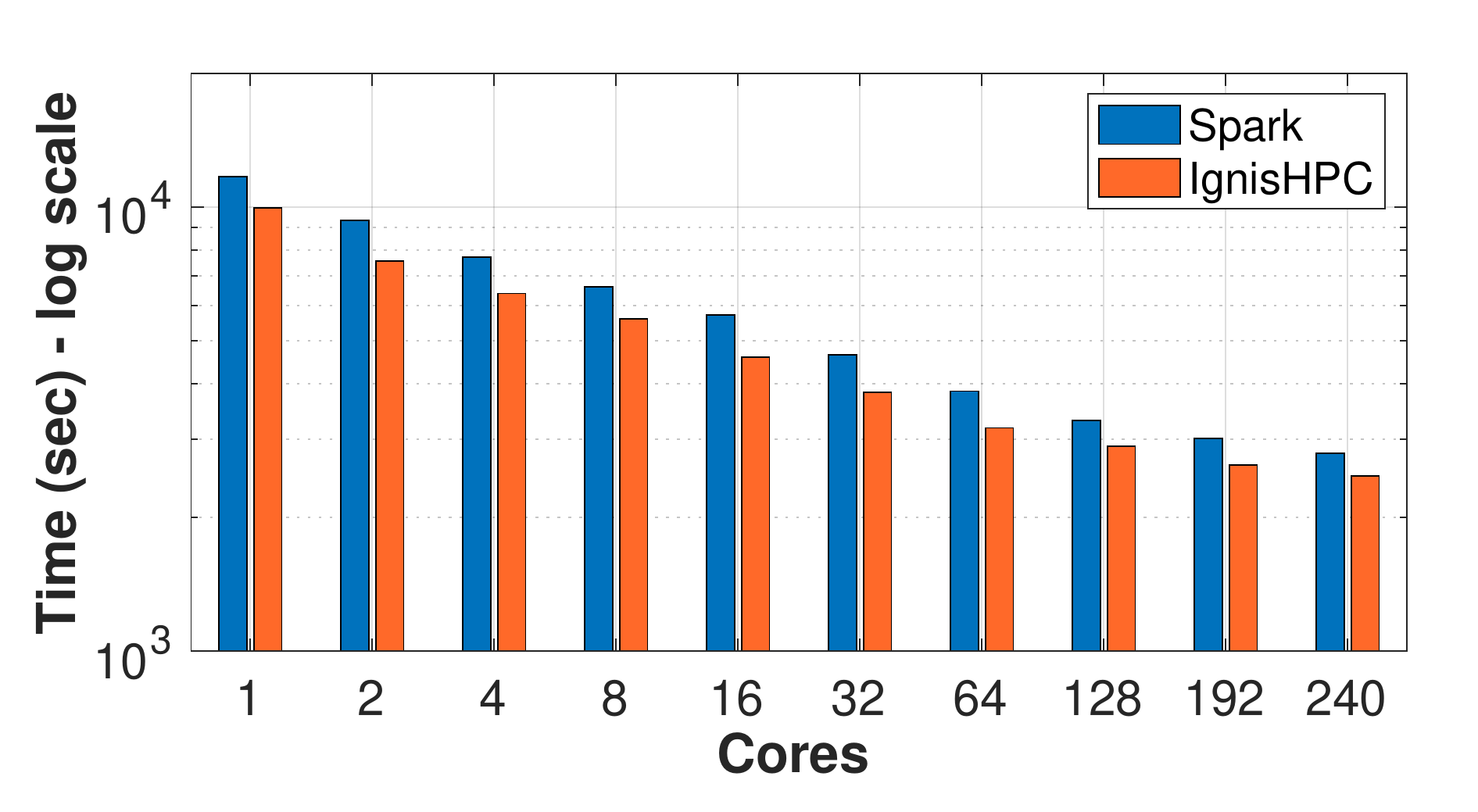}}
	\subfloat[Speedup]{\includegraphics[width=0.32\textwidth]{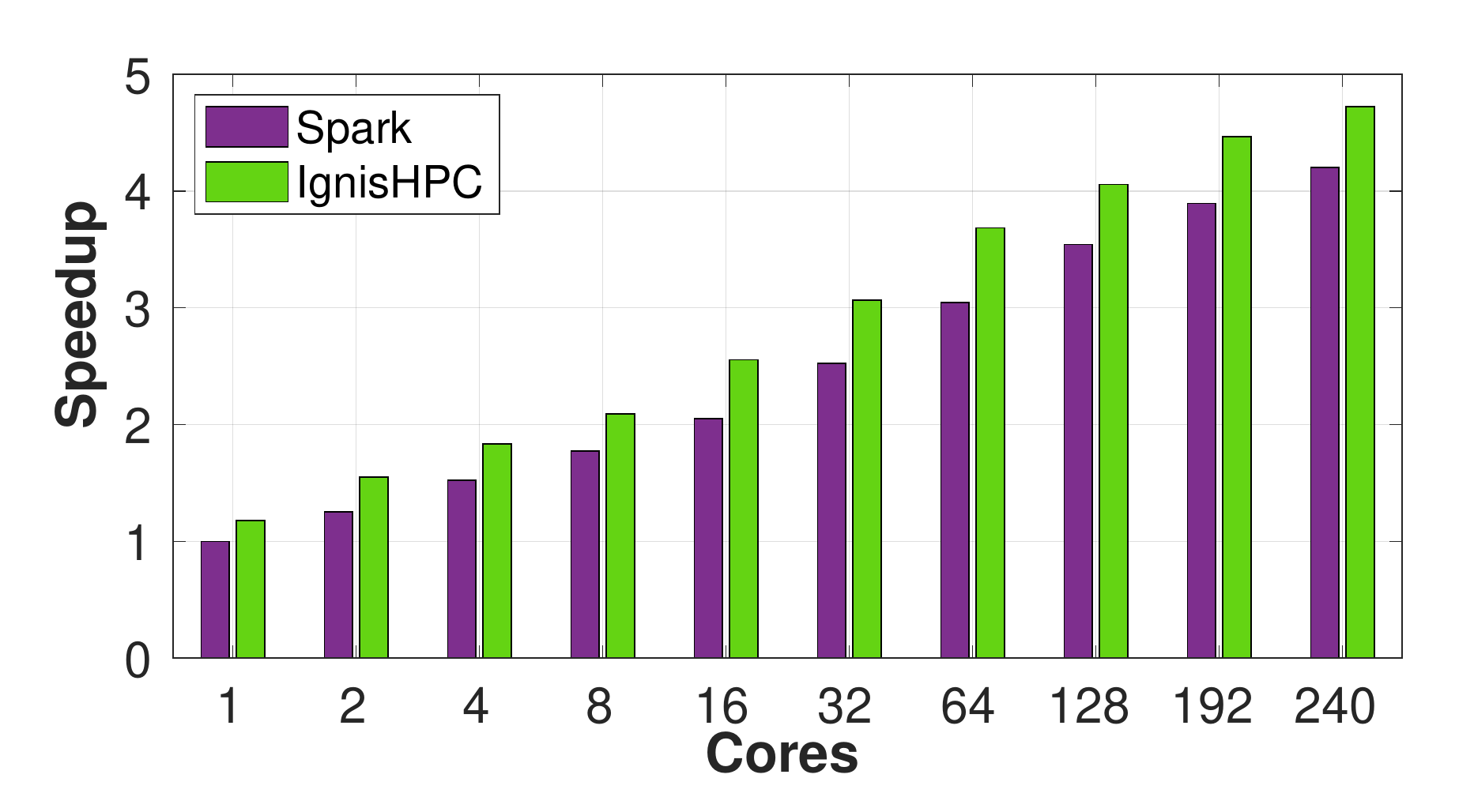}}
	\caption{Study of the scalability of \ihpc{} and Apache Spark running the Transitive Closure application.}
	\label{fig:transitive} 
\end{figure*}
\begin{table*}[t!]
\footnotesize
\centering
\begin{tabular}{lC{4cm}C{4cm}}
\hline
\rowcolor{lightgray}
{\bf Application} & {\bf No. times faster than Spark} & {\bf No. times faster than Ignis} \\ \hline
\emph{Minebench} & 3.87$\times$ [Python \& C++], 1.26$\times$ [Python] & 1.23$\times$ [Python \& C++], 1.08$\times$ [Python] \\\hline
\emph{TeraSort} & 1.76$\times$ [C++], 1.35$\times$ [Python] & --  \\\hline
\emph{K-Means} & 1.94$\times$ [Python \& C++] & 1.28$\times$ [Python \& C++]\\\hline
\emph{PageRank} & 1.10$\times$ [Python] & -- \\\hline
\emph{Transitive Closure} & 1.12$\times$ [Python] & -- \\\hline
\end{tabular}
\caption{Summary of the \ihpc{} performance results for all the Big Data applications considering the maximum number of cores and the best Ignis and Spark implementation (in case there is more than one). Between brackets the programming language/s used in the \ihpc{} implementation.}
\label{tab:bigdata_comp}
\end{table*}

In Big Data analytics many problems require processing graphs. For this reason it is essential for a Big Data framework as \ihpc{} to include primitives to support this kind of applications. In addition, since the size of the graphs to be processed is often very large, a good scalability is essential. With this goal in mind we have evaluated two well-known graph algorithms in Spark and \ihpc{}: PageRank (PR) and Transitive Closure (TC). Performance results are shown in Figures \ref{fig:pagerank} and \ref{fig:transitive}, respectively. The algorithms in Spark were implemented using GraphX~\cite{Xin13}. Note that Ignis does not support several operations that are basic for this type of applications such as \texttt{join} and \texttt{union} (see Table \ref{tab:bigdata_comp}), so it cannot be evaluated. Despite the fact that GraphX is a highly tuned API for graph processing, \ihpc{} is capable of outperforming Spark in both cases.


\begin{table*}[t!]
\footnotesize
\centering
\begin{tabular}{M{2cm}C{1.8cm}C{1.8cm}C{4.5cm}C{1.6cm}}
\hline
\rowcolor{lightgray}
& \multicolumn{2}{c}{\bf Point-to-point} & \multicolumn{2}{c}{\bf Collective}  \\ \cline{2-5} 
\rowcolor{lightgray}
\multirow{-2}{*}{\bf Application} & {\bf Blocking} & {\bf Non-blocking}    & {\bf Blocking} &  {\bf Non-blocking}    \\ \hline
\emph{LULESH} &      --       &  \texttt{Isend}, \texttt{Irecv}         &  \texttt{Allreduce}, \texttt{Barrier} &     --     \\ \hline
\emph{AMG} &  \texttt{Send}, \texttt{Recv}    &  \texttt{Isend}, \texttt{Irecv}, \texttt{Irsend} &  \texttt{Allreduce}, \texttt{Barrier}, \texttt{Bcast}, \texttt{Reduce},  \texttt{Alltoall}, \texttt{Allgather(v)}, \texttt{Gather(v)}, \texttt{Scan}, \texttt{Scatter(v)}   &     --     \\ \hline
\emph{MiniAMR}  &  \texttt{Send}, \texttt{Recv}    &  \texttt{Isend}, \texttt{Irecv} &  \texttt{Allreduce},  \texttt{Barrier}, \texttt{Bcast}, \texttt{Alltoall}                                                       &     --     \\ \hline
\emph{MiniVite} &  \texttt{Sendrecv}      &  \texttt{Isend}, \texttt{Irecv} &  \texttt{Allreduce}, \texttt{Barrier}, \texttt{Bcast}, \texttt{Reduce}, \texttt{Alltoall(v)}, \texttt{Exscan}  &  \texttt{Ialltoall}  \\ \hline
\emph{MSAProbs} &  \texttt{Send}, \texttt{Recv}    &  \texttt{Isend}, \texttt{Irecv} &  \texttt{Allreduce}, \texttt{Barrier}, \texttt{Bcast}  &     --     \\ \hline
\end{tabular}
\caption{MPI calls used for communications in the HPC applications.}
\label{tab:mpi_calls}
\end{table*}

\subsubsection{Final remarks}

Table \ref{tab:bigdata_comp} summarizes the performance gains obtained by \ihpc{} with respect to Spark and Ignis when running all the considered Big Data applications. 
Results were obtained using the maximum number of cores available and taking into account the best Ignis and Spark implementation (if there is more than one). In this way, for example, two implementations are available for Minebench, pure Python and multi-language Python-C++. In that case we used as reference for Spark the Python code, while for Ignis the best performing implementation was the multi-language one (see the values in Figures \ref{fig:py_minebench}(a) and \ref{fig:c++_minebench}(a) when using 240 cores). 

According to the results, \ihpc{} is from 1.10$\times$ to 3.87$\times$ faster than Spark. It is especially relevant the good behavior of \ihpc{} when considering multi-language applications. At the same time, \ihpc{} is a step forward with respect to Ignis in terms of performance. In this case, \ihpc{} is from 1.23$\times$ to 1.28$\times$ faster than Ignis. However, there are additional benefits. First, the memory consumption in \ihpc{} was optimized allowing multiple partitions per executor, which allows to work with extremely large datasets. That is the reason why Ignis is not able to execute TeraSort in our cluster. And second, the \ihpc{} API was extended to support, among others, graph processing algorithms such as PageRank and Transitive Closure. 

\subsection{HPC applications}
\label{sec:hpc_apps}


For many years MPI has been the dominant parallel programming model in the HPC area. As we explained in Section \ref{sec:integrating_mpi}, thanks to its architectural design, one of the most important features of \ihpc{} is its ability to execute native MPI applications within the framework just adding a few lines of code. To evaluate the benefits of our approach we are interested in two key areas: performance (with respect to the native MPI execution) and productivity (additional Source Lines Of Code - SLOC). In this way, we have selected five HPC applications coming from different scientific fields that represent a variety of MPI communication patterns. Table \ref{tab:mpi_calls} summarizes the most important MPI calls (point-to-point and collective operations) used in the applications. Note that during a specific run, an application may use only a subset of these communications. All the codes were implemented using C/C++. For some of them we have considered hybrid implementations (MPI+OpenMP) to demonstrate that is also possible to execute efficiently this type of applications in \ihpc{} without additional effort.

\begin{figure*}[t!]
	\centering
	\subfloat[Times (strong scaling)]{\includegraphics[width=0.32\textwidth]{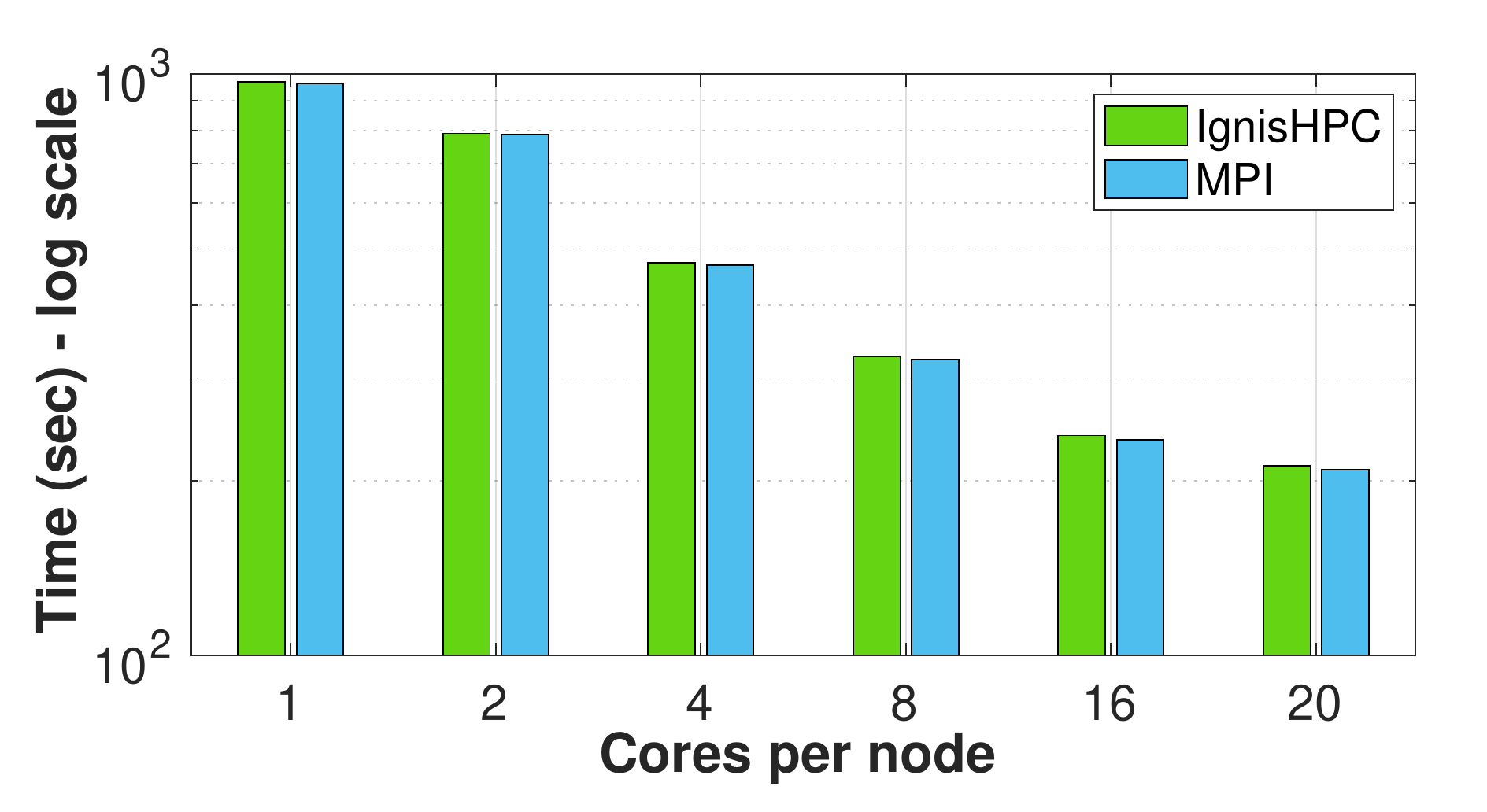}}
	\subfloat[Speedup]{\includegraphics[width=0.32\textwidth]{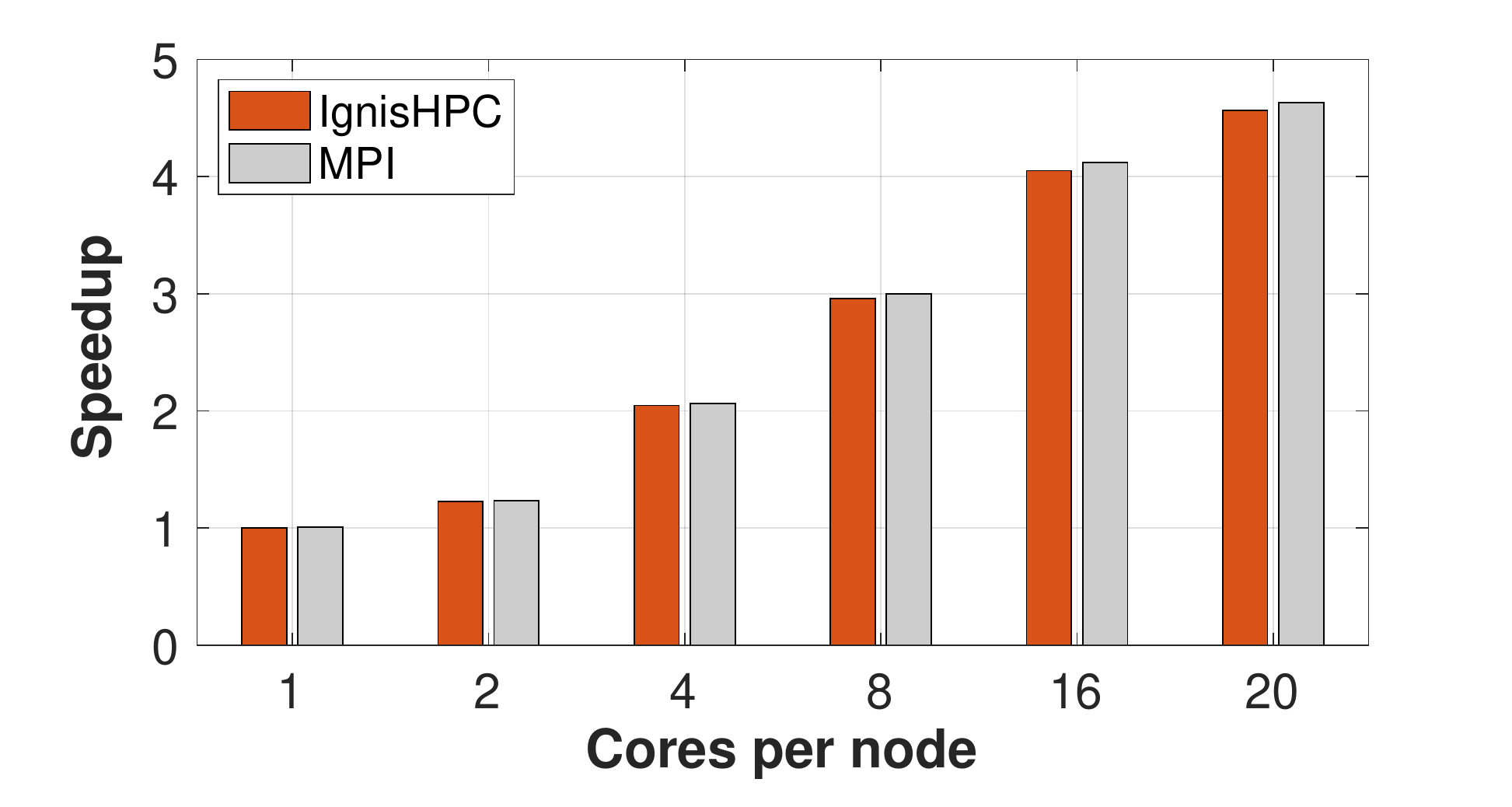}}
	\vspace{-0.25cm}
	\caption{Study of the scalability of LULESH (8 nodes).}
	\label{fig:lulesh} 
\end{figure*}

Next we provide some information about the selected HPC applications used in the experimental evaluation: 

\begin{itemize}[leftmargin=*,noitemsep]

    \item \emph{LULESH (Livermore Unstructured Lagrange Explicit Shock Hydrodynamics)}. It is a shock hydrodynamics code developed at Lawrence Livermore National Lab (LLNL)~\cite{Kar13}. It has been ported to a number of programming models: MPI, OpenMP, MPI+OpenMP, CUDA, etc. In this paper we have considered the hybrid MPI+OpenMP implementation, which uses MPI between nodes and OpenMP for cores on a node. Performance tests were run on 8 nodes with a problem size of 70$^3$ on each node, which corresponds to the most representative problem size~\cite{Kar13b}.
    
    \item \emph{AMG}. It is a parallel algebraic multigrid solver for linear systems arising from problems on unstructured grids. It is part of the Exascale Computing Project (ECP) proxy applications suite\footnote{http://proxyapps.exascaleproject.org/ecp-proxy-apps-suite} and was derived directly from the BoomerAMG~\cite{Hen02} solver. AMG is an SPMD application with about 65,000 lines of code which uses OpenMP threading within MPI tasks. Parallelism is achieved by simply subdividing the grid into logical $P\times Q\times R$ (in 3D) chunks of equal size. AMG is a highly synchronous and memory-access bound code. The scalability tests were obtained with a fixed local problem grid size per MPI process of 100$\times$100$\times$100 points.

    \item \emph{miniAMR}. It is a proxy app for adaptive mesh refinement (AMR), which is a frequently used technique for efficiently solving partial differential equations (PDEs)~\cite{Sas16}. It applies a stencil calculation on a unit cube computational domain, which is divided into blocks. This application also belongs to the ECP proxy app collection and was implemented using MPI. We used blocks with dimensions 8$\times$8$\times$8 and a maximum of 4 refinement levels. The test case we considered is that of an expanding sphere, which closely mimics an explosion. Blocks are refined along the boundary of the expanding sphere.
  
    \item \emph{miniVite}. It implements a parallel Louvain method for community detection, which is one of the most important graph kernels used in scientific and social networking applications for discovering higher order structures within a graph~\cite{Gho18}. It is also included in ECP proxy app collection. miniVite was programmed using MPI and OpenMP. As input we used a graph with 10\% of the vertices of the well-known \emph{friendster} social network graph~\cite{yan12}. It consists of 6.6M vertices and 24.2M edges.

    \item \emph{MSAProbs}. One basic step in many bioinformatics analyses is the multiple sequence alignment (MSA). MSAProbs~\cite{Liu10} is a state-of-the-art tool to compute protein MSA based on hidden Markov models. In this work we have considered its MPI+OpenMP parallel implementation~\cite{Gon16}. The input dataset $PF07085$~\cite{Mis20} used in the tests consists of 975 sequences with an average length of 512.
    
\end{itemize}

\begin{figure*}[t!]
	\centering
	\subfloat[]{\includegraphics[width=0.32\textwidth]{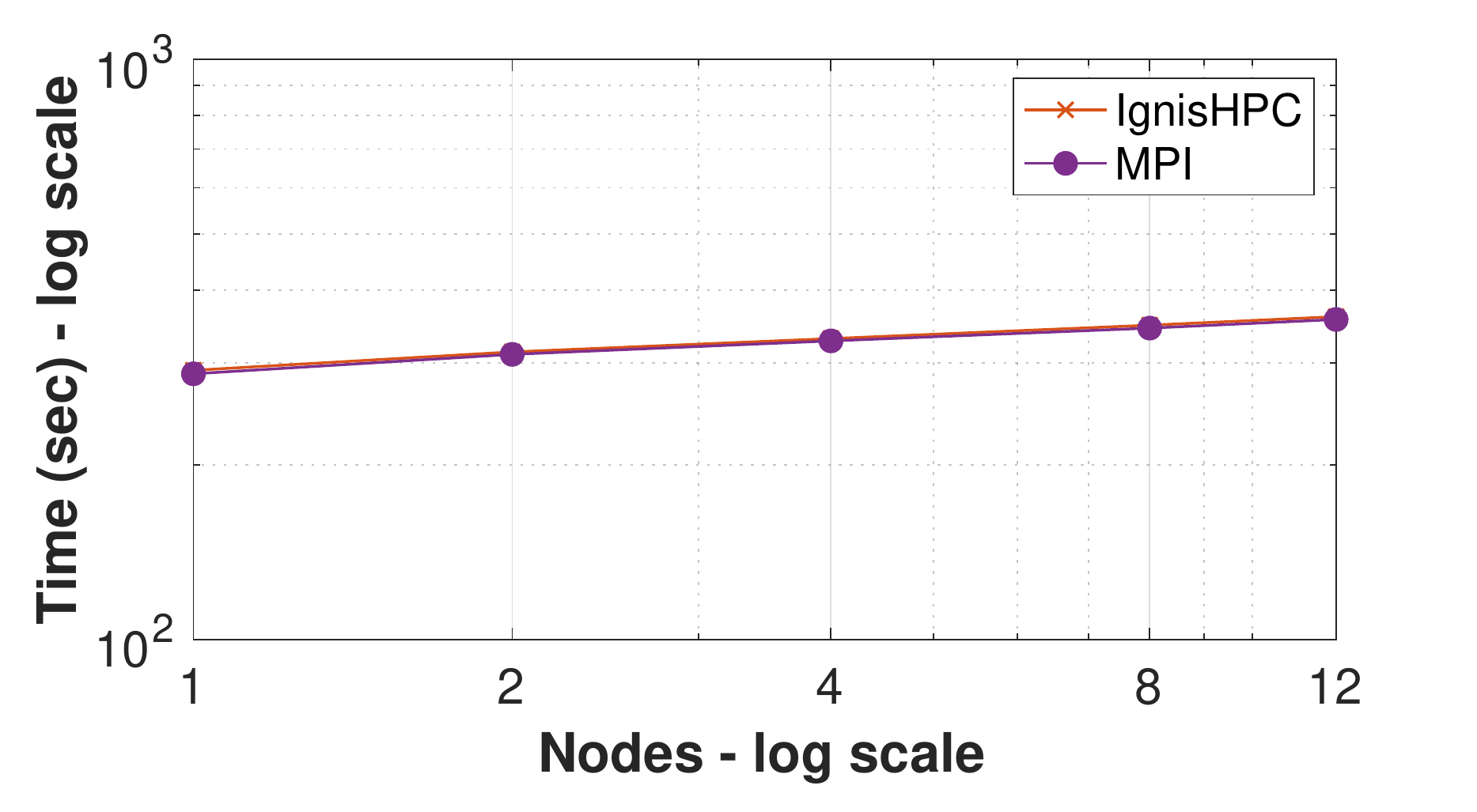}}
	\subfloat[]{\includegraphics[width=0.32\textwidth]{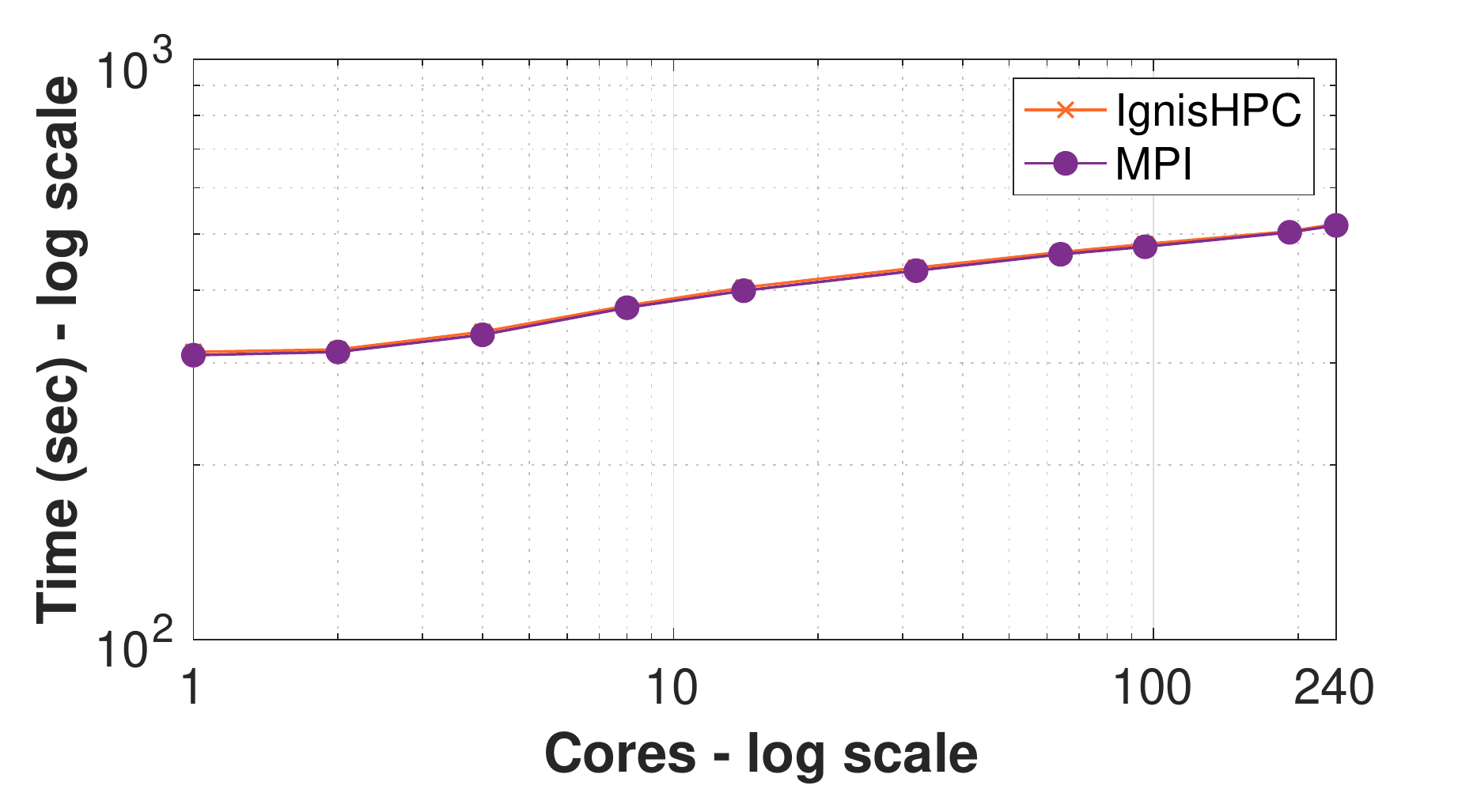}}
	\vspace{-0.25cm}
	\caption{Study of the weak scalability of AMG (20 threads/cores per node) (a) and miniAMR (b).}
	\label{fig:amg_amr} 
\end{figure*}

\subsubsection{Analysis and discussion}

Next we carry out the analysis of the execution of the MPI-based HPC applications within \ihpc{}. As mentioned previously, we will focus on two aspects. First, the performance differences between running the HPC applications on the cluster as native MPI tasks or using \ihpc{}. It is important to highlight that is out of the scope of this paper to analyze the particular behavior of each MPI application in terms of performance and scalability. This was extensively explained in the references provided in the description paragraphs of Section \ref{sec:hpc_apps}. The second key aspect is productivity. In our case we measured the source lines of code (SLOC) of the applications. 
This metric is very important since scientists will only adopt \ihpc{} to execute MPI applications if porting them requires little effort.  \\

\begin{figure*}[t!]
	\centering
	\subfloat[Times (strong scaling)]{\includegraphics[width=0.32\textwidth]{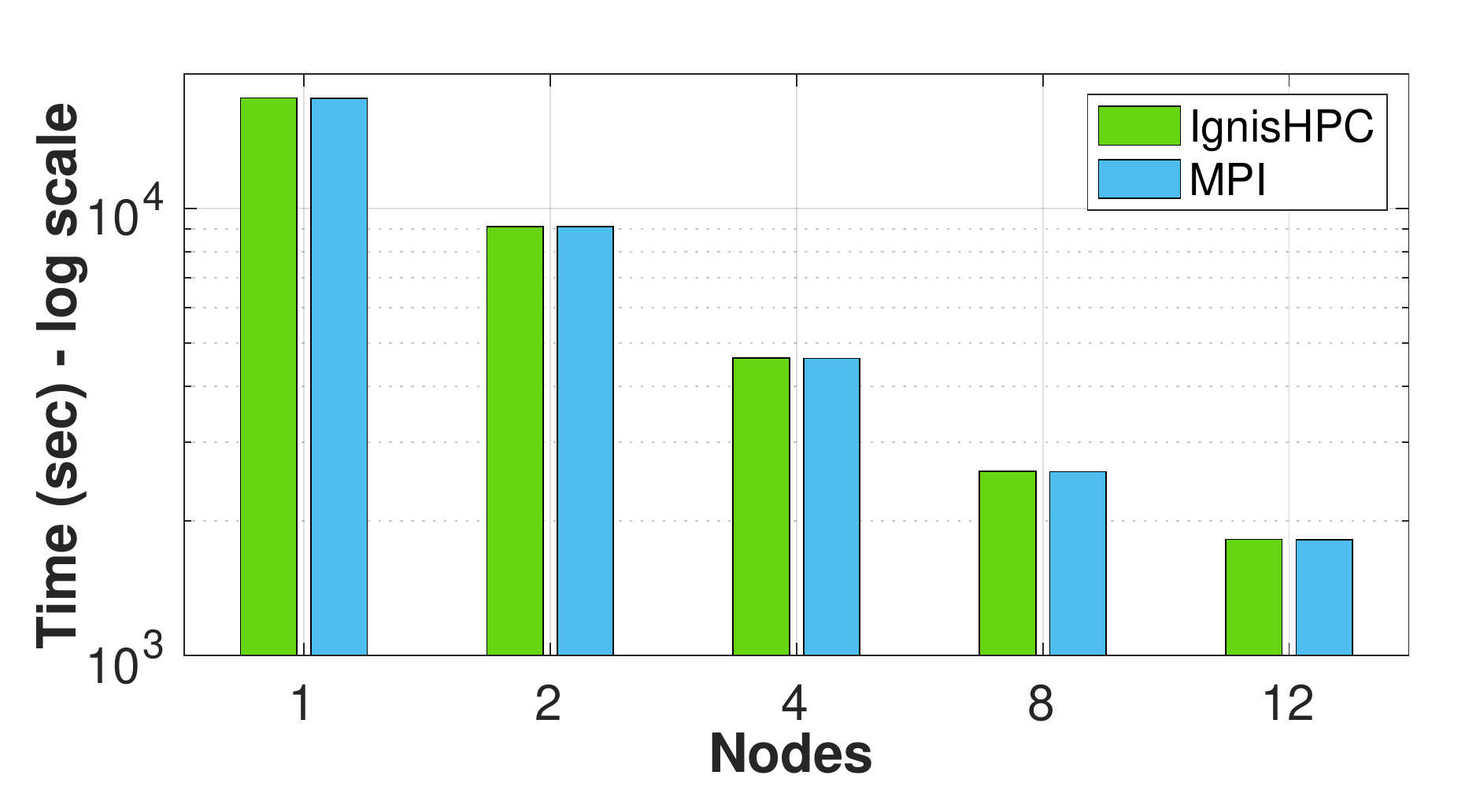}}
	\subfloat[Speedup]{\includegraphics[width=0.32\textwidth]{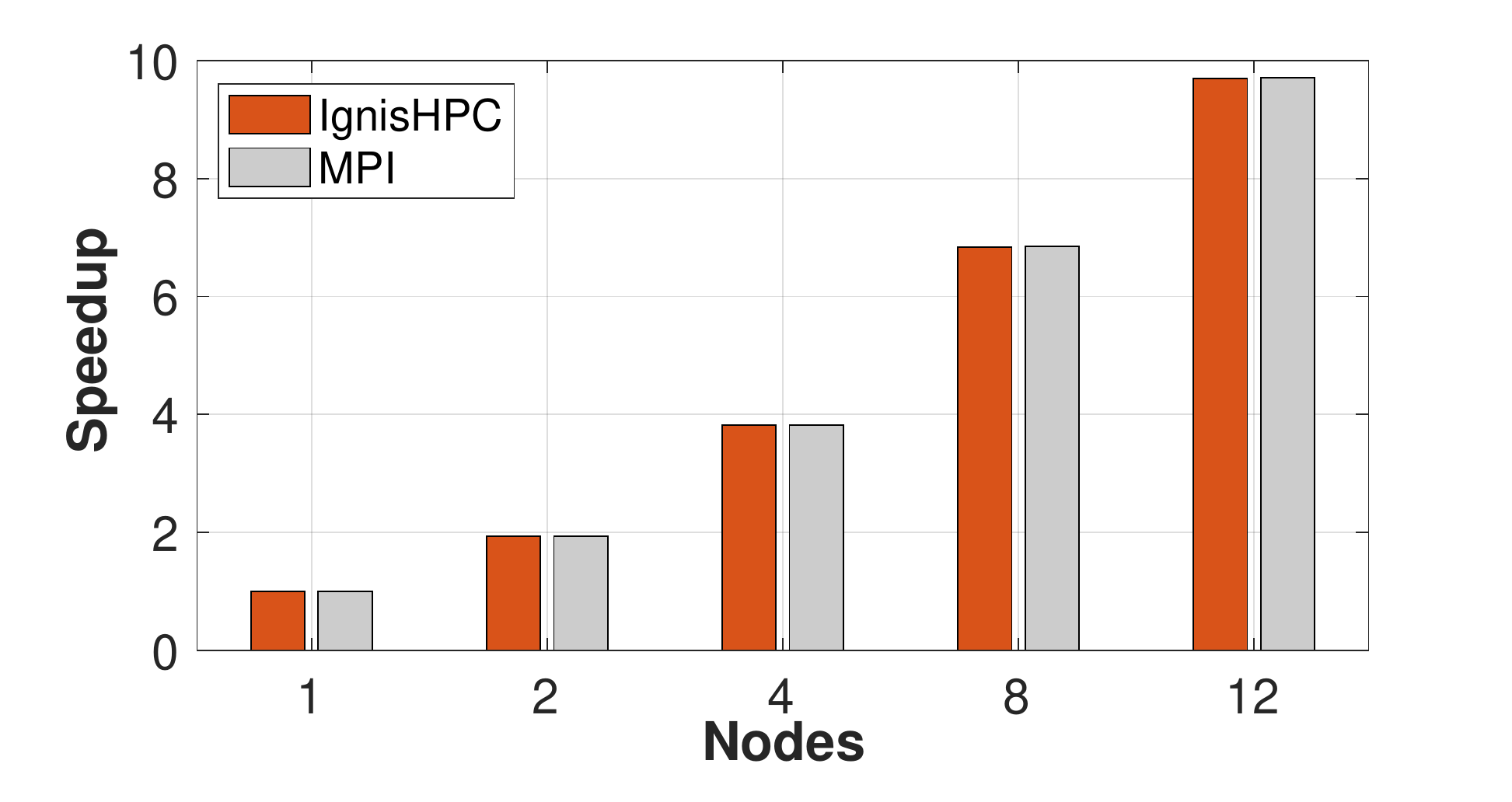}}
	\vspace{-0.25cm}
	\caption{Study of the scalability of miniVite (20 threads/cores per node).}
	\label{fig:minivite} 
\end{figure*}
\begin{figure*}[t!]
	\centering
	\subfloat[Times (strong scaling)]{\includegraphics[width=0.32\textwidth]{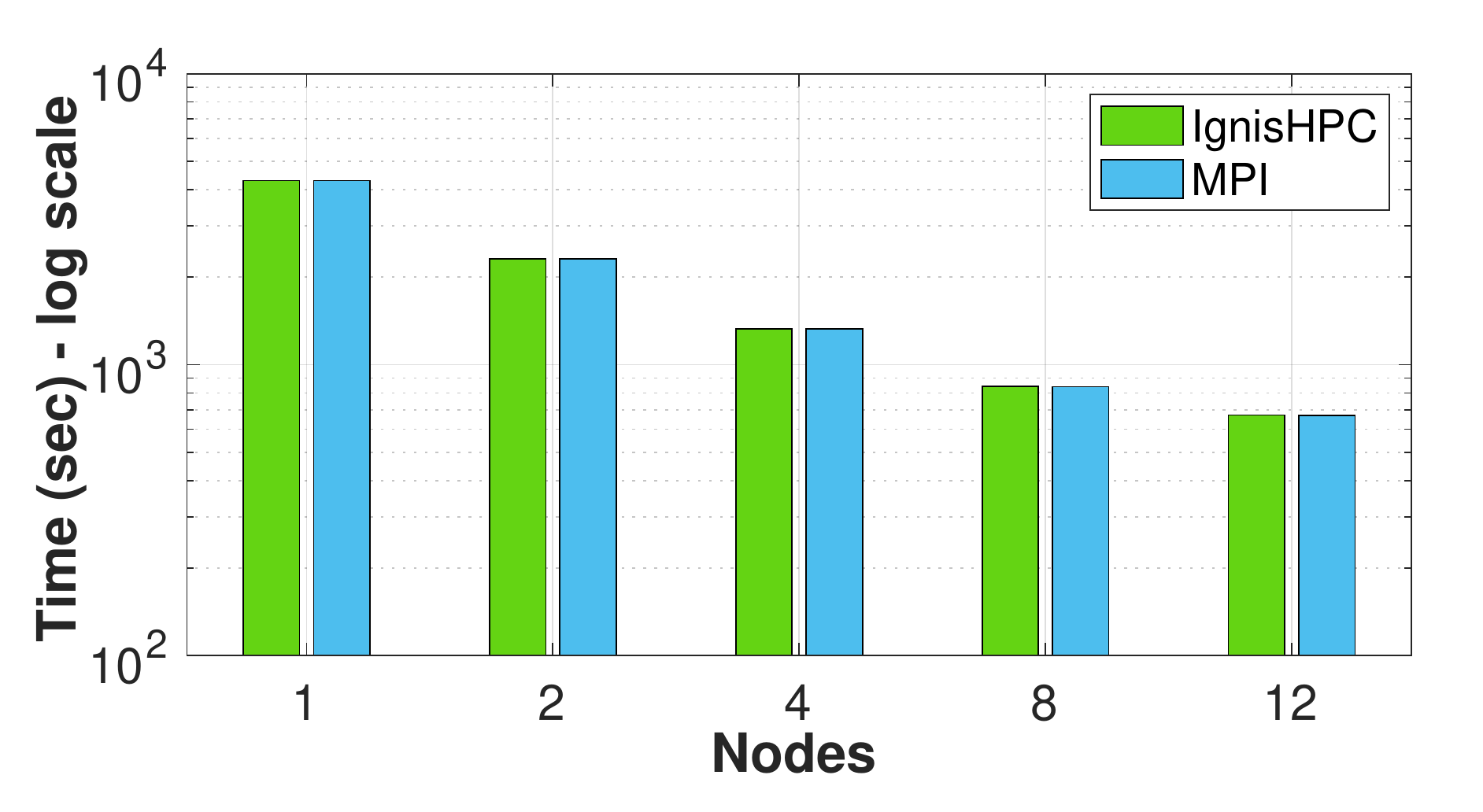}}
	\subfloat[Speedup]{\includegraphics[width=0.32\textwidth]{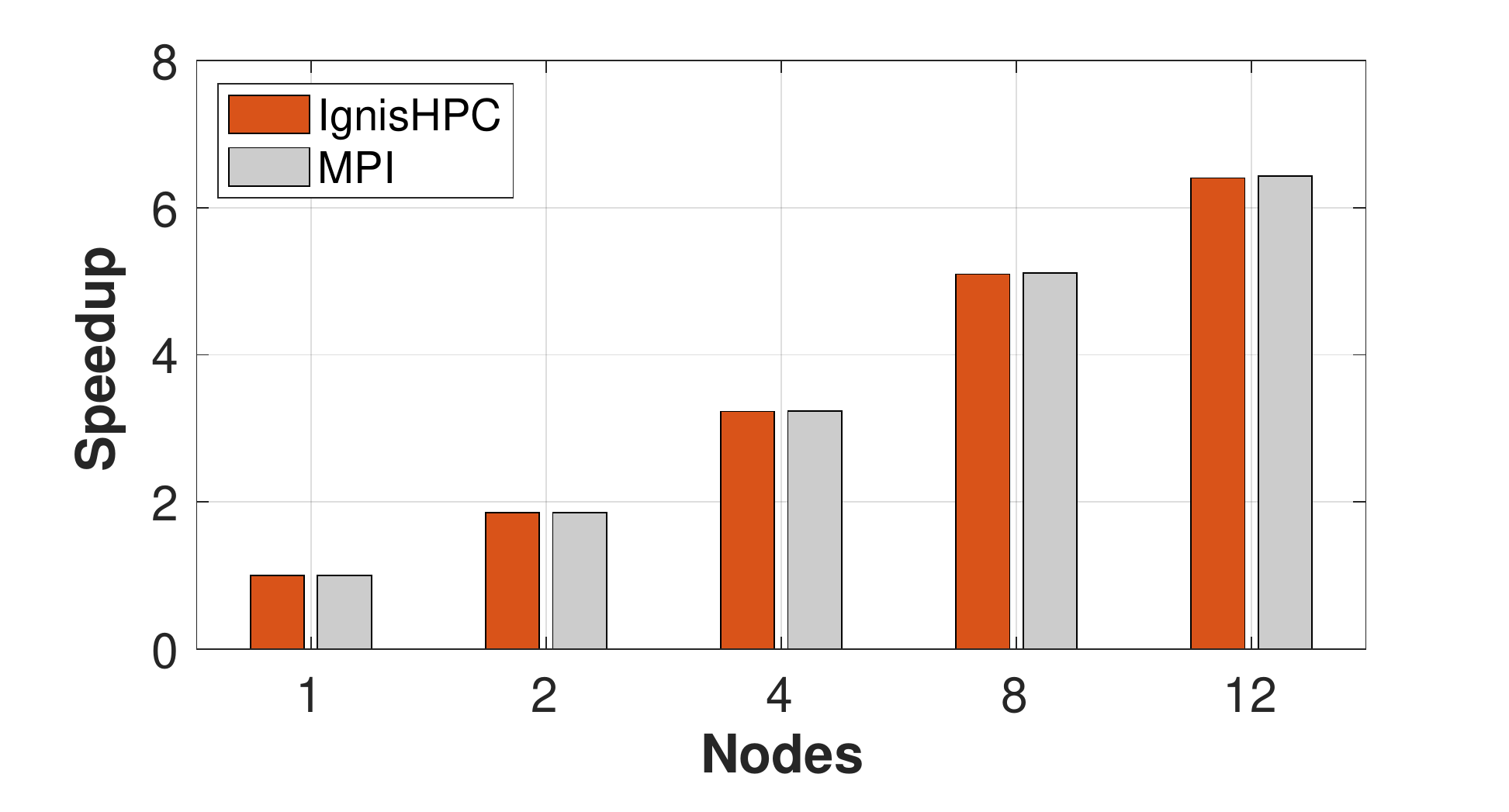}}
	\vspace{-0.25cm}
	\caption{Study of the scalability of MSAProbs (20 threads/cores per node).}
	\label{fig:msaprobs} 
\end{figure*}

\noindent \emph{Performance}. Figure \ref{fig:lulesh} shows the strong scaling results of LULESH. Note that this application uses a hybrid MPI+OpenMP approach. Performance differences between native MPI and \ihpc{} are really small, always lower than 1.7\%. Running LULESH from \ihpc{} shows the same scalability trend than the MPI native execution. 

Weak scalability tests were run to evaluate AMG (MPI + OpenMP) and miniAMR (MPI). Results are shown in Figures \ref{fig:amg_amr}(a) and \ref{fig:amg_amr}(b). In both cases also, the performance of \ihpc{} comes close to that of its counterpart, the native MPI implementation. The maximum performance difference is only about 1.4\% for both applications. In this way, for instance, the execution times of miniAMR using all the cores in the cluster were 520 and 517 seconds with native MPI and \ihpc{}, respectively. 

miniVite scalability results are displayed in Figure \ref{fig:minivite}. The behavior replicates the observations commented previously for the other HPC applications. That is, running an MPI application within the \ihpc{} framework achieves very similar performance with respect to the native execution. In this particular case, the maximum difference drops to only 0.2\%. The same scalability analysis applied to MSAProbs (Figure \ref{fig:msaprobs}) produces a maximum performance difference of 0.4\%.

So we conclude that running MPI (and MPI+OpenMP) applications from \ihpc{} is as efficient as executing them natively.  \\

\noindent \emph{Productivity}. As we explained in Section \ref{sec:integrating_mpi}, running  MPI applications in \ihpc{} requires some minimal modifications to the original source code and adding a few lines to call the application from the driver code. It is important to highlight that most of these extra lines are devoted to parsing the arguments of the MPI application. In any case, this is a very simple and repetitive code as shown in the example of Figure \ref{fig:lulesh_call}, which can be considered as \emph{boilerplate}. We measure the SLOC using SLOCcount~\cite{sloc} of each original MPI application and its counterpart adapted to \ihpc{} (see Table \ref{tab:sloc}). The number of extra lines, between brackets in the table, ranges only from 17 to 75. This demonstrates that integrating MPI applications and libraries in \ihpc{} is a straightforward process, which is very important for the HPC community since it is not necessary to port MPI codes to a new API or programming model. Therefore, \ihpc{} fulfills its design goal of unifying in a single framework the benefits of HPC and Big Data applications.  


\begin{table}[t!]
\footnotesize
\centering
\begin{tabular}{M{2.5cm}M{2.25cm}M{2.25cm}}
\hline
\rowcolor{lightgray}
{\bf Application} & {\bf SLOC MPI} & {\bf SLOC \ihpc{}} \\ \hline
\emph{LULESH} & 5,918 & 5,993 (+75) \\\hline
\emph{AMG} & 65,154 & 65,197 (+43)\\\hline
\emph{MiniAMR} & 9,958 & 9,987 (+39)\\\hline
\emph{MiniVite} & 3,264 & 3,324 (+60)\\\hline
\emph{MSAProbs}  & 6,045 & 6,062 (+17)\\
\hline
\end{tabular}
\caption{SLOC of the HPC applications.}
\label{tab:sloc}
\end{table}

\section{Related Work}
\label{sec:related}

\subsection{HPC and Containers}

HPC workloads tend to be monolithic in nature so that each component and its dependencies must be present for running or compiling the code. In addition, if an update of any component is required, all modules will be affected. For this reason the most common difficulty faced by end-users when creating and implementing scientific software is the installation and configuration of a framework with thousands of dependencies.

Containers are a good way to self-contain an application and its dependencies in a controlled environment. 
Containers do not interfere with each other and allow to be deleted or updated without leaving any trace on the physical machine. They are an alternative to virtual machines while maintaining a similar level of isolation and showing a superior performance that in some cases is almost identical to the one obtained when executing natively on a real machine~\cite{Adu15,Chu16b}.

\ihpc{} can be seen as an MPI application, so it can be efficiently executed inside containers as it was proven in several works. 
For example, running MPI applications on a containerized cluster using Docker on a cluster~\cite{Luc17} or in the Cloud~\cite{You17}, or using  Shifter~\cite{Sah18} instead. Other works deal with the orchestration of Docker containers in a HPC environment. For instance, Higgins et al.~\cite{Hig15} implemented a script based on SSH for the creation of an MPI environment inside a Docker container. Unlike \ihpc{}, this approach requires root privileges to modify the hosts configuration. Other paper introduces Scylla~\cite{Sah19}, a framework for deploying MPI jobs within Docker Containers using Apache Mesos. As explained in Section \ref{sec:background}, Apache Mesos requires an orchestration framework such as Marathon or Singularity to be used. However, the authors, instead of considering a third party framework, implemented an ad-hoc solution. Their approach also requires root privileges. We must highlight that \ihpc{} supports all the functionalities included in Scylla without needing root permissions and is not limited to work with Apache Mesos. 

\subsection{Spark and HPC applications}

As a general-purpose framework, Spark has been widely used for many scientific applications and algorithms. However, there are examples from different areas such as linear algebra~\cite{git16b}, genomics~\cite{Abu20} or even data science~\cite{Sax20} where Spark does not obtain the expected performance. 

One way to approach Big Data and HPC worlds is trying to boost the performance of well-established Big Data technologies when running on HPC systems. For example, taking advantage of the Infiniband fast interconnection network~\cite{Lu13} or the standard HPC programming models such as MPI. We are especially interested in those works that opt for the latter approach. For instance, Anderson et al.~\cite{And17} try to combine Spark and MPI. They offload computations to an MPI environment from within Spark in such a way that Spark and MPI tasks run at the same time, using a socket-based implementation for efficient data exchange between processes. In their approach the results of the MPI processing are copied back to persistent storage (HDFS), and then into Spark for further processing. As a consequence, for those applications that require few iterations and/or less work per iteration, there is a degradation in the performance. However, their approach shows a good behavior with several graph and machine learning applications that do not require a lot of data movement between Spark and MPI environments. 

A similar solution can be found in~\cite{Git18}. They introduce Alchemist, a TCP socket-based implementation for inter-process communication between Spark and MPI. Alchemist was designed to call MPI-based libraries from Spark using the Scala programming language. Like previous work, due to the use of two type of tasks for MPI and Spark, it is always necessary to keep two copies of the same data. In addition, moving data between Spark and MPI processes is costly since TCP sockets are often a slower alternative with respect to shared memory. Therefore, this approach is also limited to computationally-intensive applications for which the cost associated with data transfers is negligible when compared to the overhead that would have been incurred by Spark.

Finally, there are several related works of the same research group that make a better integration between Spark and HPC technologies. In \cite{Mal16}, Spark and MPI tasks share the same process, which removes the overhead caused by the data transfers. Python is used as programming language since Spark and MPI has an interface for this language. However, as we explained in~\cite{Pin20}, Python is not natively supported by Spark which causes an important degradation in its overall performance. \ihpc{} overcomes that limitation using native executors for each supported programming language.

In their next works, the authors introduce Spark-MPI~\cite{Mal17,Mal18}, a hybrid platform that combines Spark and MPI taking advantage of the MPI Exascale Process Management Interface (PMIx). A Spark-MPI application consists of a driver launched by Spark and a set of processes launched by MPI. These MPI processes connect to the Spark driver as workers and they will be able to execute both RDD and MPI functions. Note that using MPI routines in Spark-MPI only makes sense when the data is processed using \texttt{mapPartitions}. That is the only way that Spark provides to work on a complete partition instead of on each element of the partition. The authors showed the benefits of Spark-MPI with deep learning algorithms and ptychographic and tomographic applications. However, Spark-MPI has several limitations that we detailed next:

\begin{itemize}[noitemsep]
    \item Spark-MPI requires a hybrid environment for Spark and MPI, which will be configured with their respective resource managers. For example, Mesos or Yarn for Spark and Hydra or Slurm for MPI. It is hard to find a system configured this way since resource managers cannot shared the available hardware resources. Moreover, a Spark-MPI job should queue for both Spark and MPI queuing systems, and the requested resources may not be granted at the same time. On the other hand, \ihpc{} is a single application so the previous problems do not apply.
    \item Due to its particular architecture and how Spark executors are launched, Spark-MPI loses the fault tolerance system provided by Spark. As a consequence, after any failure in the executors, all the job is lost. On the contrary, as we demonstrated in~\cite{Pin20}, Ignis and \ihpc{} are able to recover after a failure of a cluster node or some of the executors. If some data is lost, \ihpc{} has enough information about how it was derived in such a way that only those operations needed to recompute the corresponding portion of data are performed.
    \item Spark-MPI is limited to access one partition at the same time per executor. Partition size is restricted to a maximum of 2 GB, which is related to the use of JVMs in Spark. Therefore, additional executors should be created in case more data is necessary, degrading the overall I/O performance. This restriction only applies to \ihpc{} for Java applications, but not for Python and C/C++.    
    \item Although the execution of pure MPI codes in Spark-MPI is discussed, the vast majority of MPI libraries and applications are implemented in C/C++ for which Spark does not have native support. Spark uses system pipes to share data outside the JVM in order to run non-JVM codes, which introduces an additional overhead that negatively affects performance. So the expected performance of running MPI applications in the Spark-MPI platform would be low. However, as we demonstrated in this paper, running MPI applications in \ihpc{} achieves very similar performance with respect to its native execution.
\end{itemize}

\section{Conclusions}
\label{sec:conclusions}

In this work we have introduced a new computing framework named \ihpc{}\footnote{It is publicly available at \url{https://github.com/ignishpc}} to fill the gap between Big Data and HPC languages and programming models. \ihpc{} supports the combination of JVM and non-JVM-based languages in the same application (currently, Java, Python and C/C++). It was designed to take advantage of MPI for communications, which allows the framework to execute efficiently MPI applications and libraries. As a consequence, the MPI-based HPC scientific applications do not have to be ported to a new API or programming model. Moreover, it is possible to combine in the same multi-language code HPC tasks (using MPI) with Big Data tasks (using MapReduce operations).

The experimental evaluation demonstrated the benefits of our proposal in terms of performance with respect to the \emph{de-facto} standard for Big  Data processing, Spark, and our first prototype of multi-language framework, Ignis. In particular, \ihpc{} is from 1.1$\times$ to 3.9$\times$ faster than Spark, and about 1.2$\times$ faster than Ignis. In the same way, we observed that running MPI and MPI+OpenMP  applications in \ihpc{} is as efficient as executing them natively. Therefore, thanks to \ihpc{} we are merging both Big Data and HPC software ecosystems in just one execution environment. 

\


\bibliographystyle{elsarticle-num} 
\bibliography{ms}

\begin{thebibliography}{10}
\expandafter\ifx\csname url\endcsname\relax
  \def\url#1{\texttt{#1}}\fi
\expandafter\ifx\csname urlprefix\endcsname\relax\def\urlprefix{URL }\fi
\expandafter\ifx\csname href\endcsname\relax
  \def\href#1#2{#2} \def\path#1{#1}\fi

\bibitem{Hel20}
S.~Heldens, \emph{et al.}, {The Landscape of Exascale Research: A Data-Driven
  Literature Analysis}, ACM Comput. Surv. 53~(2) (2020).

\bibitem{Whi15}
T.~White, Hadoop: The Definitive Guide, 4th Edition, O'Reilly Media, Inc.,
  2015.

\bibitem{Zah10}
M.~Zaharia, M.~Chowdhury, M.~J. Franklin, S.~Shenker, I.~Stoica, {Spark:
  Cluster Computing with Working Sets}, in: Proc. of the 2nd USENIX Conf. on
  Hot Topics in Cloud Computing (HotCloud), 2010, pp. 10--10.

\bibitem{Asc18}
M.~Asch, \emph{et al.}, {Big data and extreme-scale computing: Pathways to
  Convergence-Toward a shaping strategy for a future software and data
  ecosystem for scientific inquiry}, The International Journal of High
  Performance Computing Applications 32~(4) (2018) 435--479.

\bibitem{dea04}
J.~Dean, S.~Ghemawat, {MapReduce: Simplified Data Processing on Large
  Clusters}, in: Symposium on Operating System Design and Implementation, 2004,
  pp. 10--10.

\bibitem{Ding2011}
M.~Ding, \emph{et al.}, {More Convenient More Overhead: The Performance
  Evaluation of Hadoop Streaming}, in: Proc. of the ACM Symposium on Research
  in Applied Computation, 2011, pp. 307--313.

\bibitem{Jython}
{Jython}, \url{http://www.jython.org/}, [Online; accessed April, 2019].

\bibitem{Pin20}
C.~Piñeiro, R.~Martínez-Castaño, J.~C. Pichel, {Ignis: An efficient and
  scalable multi-language Big Data framework}, Future Generation Computer
  Systems 105 (2020) 705--716.

\bibitem{mpich}
{MPICH}, \url{https://www.mpich.org}, [Online; accessed October, 2021].

\bibitem{openmpi}
{Open-MPI}, {\tt https://www.open-mpi.org/}, [Online; accessed October, 2021].

\bibitem{Hin11}
B.~Hindman, \emph{et al.}, {Mesos: A Platform for Fine-Grained Resource Sharing
  in the Data Center}, in: Proc. of the 8th USENIX Conf. on Networked Systems
  Design and Implementation, 2011, p. 295–308.

\bibitem{nomad}
{HashiCorp}, Nomad: workload orchestration made easy,
  \url{https://www.nomadproject.io/}, [Online; accessed October, 2021].

\bibitem{mer14}
D.~Merkel, Docker: lightweight linux containers for consistent development and
  deployment, Linux journal 2014~(239) (2014) 2.

\bibitem{marathon}
{Apache Marathon}, {\tt https://mesosphere.github.io/marathon/}.

\bibitem{singularity}
{Apache Singularity}, {\tt https://getsingularity.com/}.

\bibitem{Gopal2014}
J.~T. Kukunas, V.~Gopal, J.~Guilford, S.~Gulley, A.~van~de Ven, W.~Feghali,
  {High Performance {ZLIB} Compression on Intel Architecture Processors}, Tech.
  rep., Intel (2014).

\bibitem{Zah12}
M.~Zaharia, \emph{et al.}, {Resilient Distributed Datasets: A Fault-tolerant
  Abstraction for In-memory Cluster Computing}, in: Proceedings of the 9th
  USENIX Conference on Networked Systems Design and Implementation, USENIX
  Association, 2012, pp. 2--2.

\bibitem{Bay17}
M.~de~Bayser, R.~Cerqueira, {Integrating MPI with Docker for HPC}, in: IEEE
  Int. Conference on Cloud Engineering (IC2E), 2017, pp. 259--265.

\bibitem{Kar13}
I.~Karlin, \emph{et al.}, {Exploring Traditional and Emerging Parallel
  Programming Models Using a Proxy Application}, in: 27th Int. Symposium on
  Parallel and Distributed Processing, 2013, pp. 919--932.

\bibitem{Vav13}
V.~K. Vavilapalli, \emph{et al.}, {Apache Hadoop YARN: Yet Another Resource
  Negotiator}, in: Proc. of the 4th Annual Symposium on Cloud Computing, ACM,
  2013, pp. 5:1--5:16.

\bibitem{Bitcoin2008}
S.~Nakamoto, \href{http://bitcoin.org/bitcoin.pdf}{{Bitcoin: A Peer-to-Peer
  Electronic Cash System}} (2008).
\newline\urlprefix\url{http://bitcoin.org/bitcoin.pdf}

\bibitem{Merkle}
R.~C. Merkle, Protocols for public key cryptosystems, in: IEEE Symposium on
  Security and Privacy, 1980, pp. 122--122.

\bibitem{Li1993}
X.~Li, P.~Lu, J.~Schaeffer, J.~Shillington, P.~S. Wong, H.~Shi, {On the
  Versatility of Parallel Sorting by Regular Sampling}, Parallel Computing 19
  (1993) 1079--1103.

\bibitem{Chu09}
T.-S. Chua, J.~Tang, R.~Hong, H.~Li, Z.~Luo, Y.~Zheng, {NUS-WIDE: A Real-world
  Web Image Database from National University of Singapore}, in: Proc. of the
  ACM Int. Conference on Image and Video Retrieval (CIVR), 2009, pp.
  48:1--48:9.

\bibitem{snap}
J.~Leskovec, A.~Krevl, {SNAP Datasets: Stanford Large Network Dataset
  Collection}, \url{http://snap.stanford.edu/data} (2014).

\bibitem{Men16}
X.~Meng, \emph{et al.}, {MLlib: Machine Learning in Apache Spark}, The Journal
  of Machine Learning Research 17~(1) (2016) 1235--1241.

\bibitem{Xin13}
R.~S. Xin, J.~E. Gonzalez, M.~J. Franklin, I.~Stoica, {GraphX: A Resilient
  Distributed Graph System on Spark}, in: 1st International Workshop on Graph
  Data Management Experiences and Systems, ACM, 2013.

\bibitem{Kar13b}
I.~Karlin, J.~McGraw, J.~Keasler, B.~Still, {Tuning the LULESH Mini-app for
  Current and Future Hardware}, Tech. rep. (2013).

\bibitem{Hen02}
V.~E. Henson, U.~M. Yang, {BoomerAMG: A parallel algebraic multigrid solver and
  preconditioner}, Applied Numerical Mathematics 41~(1) (2002) 155--177.

\bibitem{Sas16}
A.~Sasidharan, M.~Snir, {MiniAMR - A miniapp for Adaptive Mesh Refinement},
  Tech. rep. (2016).

\bibitem{Gho18}
S.~Ghosh, M.~Halappanavar, A.~Tumeo, A.~Kalyanaraman, A.~H. Gebremedhin,
  {MiniVite: A Graph Analytics Benchmarking Tool for Massively Parallel
  Systems}, in: IEEE/ACM Performance Modeling, Benchmarking and Simulation of
  High Performance Computer Systems (PMBS), 2018, pp. 51--56.

\bibitem{yan12}
J.~Yang, J.~Leskovec, {Defining and Evaluating Network Communities based on
  Ground-truth} (2012).
\newblock \href {http://arxiv.org/abs/1205.6233} {\path{arXiv:1205.6233}}.

\bibitem{Liu10}
Y.~Liu, B.~Schmidt, D.~L. Maskell, {MSAProbs: multiple sequence alignment based
  on pair hidden Markov models and partition function posterior probabilities},
  Bioinformatics 26~(16) (2010) 1958--1964.

\bibitem{Gon16}
J.~González-Domínguez, Y.~Liu, J.~Touriño, B.~Schmidt, {MSAProbs-MPI:
  parallel multiple sequence aligner for distributed-memory systems},
  Bioinformatics 32~(24) (2016) 3826--3828.

\bibitem{Mis20}
J.~Mistry, \emph{et al.}, {{Pfam: The protein families database in 2021}},
  Nucleic Acids Research 49~(D1) (2020) D412--D419.

\bibitem{sloc}
D.~Wheeler, {SLOCCount}, \url{http://www.dwheeler.com/sloccount}, [Online;
  accessed November, 2021].

\bibitem{Adu15}
T.~Adufu, J.~Choi, Y.~Kim, Is container-based technology a winner for high
  performance scientific applications?, in: 17th Asia-Pacific Network
  Operations and Management Symp. (APNOMS), 2015, pp. 507--510.

\bibitem{Chu16b}
M.~T. Chung, N.~Quang-Hung, M.-T. Nguyen, N.~Thoai, {Using Docker in high
  performance computing applications}, in: IEEE 6th Int. Conference on
  Communications and Electronics (ICCE), 2016, pp. 52--57.

\bibitem{Luc17}
L.~Benedicic, F.~A. Cruz, A.~Madonna, K.~Mariotti, {Portable, high-performance
  containers for HPC} (2017).
\newblock \href {http://arxiv.org/abs/1704.03383} {\path{arXiv:1704.03383}}.

\bibitem{You17}
A.~J. Younge, K.~Pedretti, R.~E. Grant, R.~Brightwell, {A Tale of Two Systems:
  Using Containers to Deploy HPC Applications on Supercomputers and Clouds},
  in: IEEE Int. Conference on Cloud Computing Technology and Science
  (CloudCom), 2017, pp. 74--81.

\bibitem{Sah18}
P.~Saha, A.~Beltre, P.~Uminski, M.~Govindaraju, {Evaluation of Docker
  Containers for Scientific Workloads in the Cloud}, in: Proc. of the Practice
  and Experience on Advanced Research Computing, 2018.

\bibitem{Hig15}
J.~Higgins, V.~Holmes, C.~Venters, {Orchestrating Docker Containers in the HPC
  Environment}, in: HPC. Lecture Notes in Computer Science, vol 9137., Springer
  Int. Publishing, 2015, pp. 506--513.

\bibitem{Sah19}
P.~Saha, A.~Beltre, M.~Govindaraju, {Scylla: a Mesos Framework for Container
  Based MPI Jobs}, CoRR abs/1905.08386 (2019).
\newblock \href {http://arxiv.org/abs/1905.08386} {\path{arXiv:1905.08386}}.

\bibitem{git16b}
A.~Gittens, \emph{et al.}, {Matrix factorizations at scale: A comparison of
  scientific data analytics in Spark and C+MPI using three case studies}, in:
  IEEE Int. Conf. on Big Data, 2016, pp. 204--213.

\bibitem{Abu20}
J.~M. Abuín, N.~Lopes, L.~Ferreira, T.~F. Pena, B.~Schmidt, {Big Data in
  metagenomics: Apache Spark vs MPI}, Plos One 15~(10) (2020) 1--20.

\bibitem{Sax20}
M.~Saxena, S.~Jha, S.~Khan, J.~Rodgers, P.~Lindner, E.~Gabriel, {Comparison of
  MPI and Spark for Data Science Applications}, in: IEEE Int. Parallel and
  Distributed Processing Symposium Workshops (IPDPSW), 2020, pp. 682--690.

\bibitem{Lu13}
X.~{Lu}, N.~S. {Islam}, M.~{Wasi-Ur-Rahman}, J.~{Jose}, H.~{Subramoni},
  H.~{Wang}, D.~K. {Panda}, {High-Performance Design of Hadoop RPC with RDMA
  over InfiniBand}, in: 42nd Int. Conference on Parallel Processing, 2013, pp.
  641--650.

\bibitem{And17}
M.~Anderson, \emph{et al.}, {Bridging the Gap between HPC and Big Data
  Frameworks}, Proc. VLDB Endow. 10~(8) (2017) 901–912.

\bibitem{Git18}
A.~Gittens, \emph{et al.}, {Accelerating Large-Scale Data Analysis by
  Offloading to High-Performance Computing Libraries Using Alchemist}, in:
  Proc. of the 24th ACM SIGKDD Int. Conference on Knowledge Discovery \& Data
  Mining, 2018, p. 293–301.

\bibitem{Mal16}
N.~Malitsky, {Bringing the HPC reconstruction algorithms to Big Data
  platforms}, in: NY Scientific Data Summit (NYSDS), 2016, pp. 1--8.

\bibitem{Mal17}
N.~{Malitsky}, \emph{et al.}, {Building near-real-time processing pipelines
  with the Spark-MPI platform}, in: NY Scientific Data Summit (NYSDS), 2017,
  pp. 1--8.

\bibitem{Mal18}
N.~Malitsky, R.~Castain, M.~Cowan, {Spark-MPI: Approaching the Fifth Paradigm
  of Cognitive Applications} (2018).
\newblock \href {http://arxiv.org/abs/1806.01110} {\path{arXiv:1806.01110}}.

\end{thebibliography}





\end{document}